\documentclass{aa}

\usepackage{graphicx}
\usepackage[usenames,dvipsnames]{xcolor}
\usepackage{amsmath}
\usepackage{multirow}
\usepackage{float}
\usepackage{cancel}
\usepackage{txfonts}
\usepackage[colorlinks=true,allcolors=blue]{hyperref}
\usepackage{multirow}
\usepackage{url}
\usepackage{cancel}
\usepackage[normalem]{ulem}

\usepackage{comment}

\newcommand{\elev}{^}
\newcommand{\MD}{\hyperlink{cite.MunozDarias2016}{MD16}}

\newcommand{\MS}{\hyperlink{cite.MataSanchez2018}{MS18}}

\titlerunning{Modelling the Nebular Phase of V404~Cyg} 

\begin{document}

   \title{The Physics of the Nebular Phase in Black Hole X-ray Binaries: Modelling V404~Cygni}

   \author{A. Ambrifi\inst{1,}\inst{2}\fnmsep\thanks{ambrifi.astronomy@gmail.com}
          \and
          T. Muñoz-Darias\inst{1,}\inst{2}
          \and
          J. A. Fernández-Ontiveros\inst{3}
          \and
          J. Casares\inst{1,}\inst{2}
          \and
          D. Mata Sánchez\inst{1,}\inst{2}
          \and
          J. H. Matthews\inst{4}
          }

   \institute{Instituto de Astrofísica de Canarias, E-38205 La Laguna, Tenerife, Spain
         \and
             Departamento de astrofísica, Universidad de La Laguna, E-38206 La Laguna, Tenerife, Spain  
        \and
            Centro de Estudios de Física del Cosmos de Aragón (CEFCA), Plaza San Juan 1, 44001 Teruel, Spain
        \and
            Astrophysics Subdepartment, Department of Physics, University of Oxford, Keble Road, Oxford, OX13RH, UK
            }

   \date{Received Month day, year; accepted Month day, year}

\abstract{
Some of the most extreme galactic black hole transients exhibit a nebular phase at the end of their outbursts. This stage was first identified in the 2015 outburst of V404~Cygni, when it dominated the optical spectrum for \textasciitilde4 days. It is characterised by exceptionally strong optical emission lines, together with a significant increase in the Balmer decrement. The emission features show broad wings reaching velocities consistent with outflow speeds derived from prominent P-Cygni profiles that preceded the nebular phase. They are thus interpreted as arising from previously launched ejecta that expand and recombine as X-ray irradiation declines.

In this work, we modelled this phase using the photoionisation code \textsc{Cloudy}. Adopting simple assumptions for the irradiating continuum and emitting gas, we quantitatively reproduced key observables, such as the Balmer decrement and the H$\alpha$ equivalent width, and qualitatively reproduced the observed spectra. Our simulations show that the Balmer decrement remains roughly stable within the observed range across orders-of-magnitude changes in the density and ionisation parameter, spanning $n_\mathrm{H} \sim 10^{10.5}$–$10^{12.5}$\,cm$^{-3}$ and $-1.5 \lesssim \log\xi \lesssim 1$. This supports the scenario in which the nebular phase can persist through moderate ejecta expansion and cooling, consistent with observations. Finally, we used different approaches to constrain the general properties of the nebular phase, which we propose is produced by clumpy, partially ionised ejecta whose total mass likely exceeds the mass accreted during the outburst.}


   \maketitle

\section{Introduction}
\label{sec:intro}
Black hole X-ray binaries (BHXBs)  are binary systems in which a stellar-mass black hole accretes matter from a companion star via an accretion disc. These objects spend most of their time in a dim, quiescent state but sporadically exhibit energetic outbursts, characterised by a several-orders-of-magnitude increase in their X-ray luminosity. During these events, blueshifted absorptions are often observed in their X-ray spectra (e.g. \citealt{Neilsen2009, Ueda2009, Ponti2012, DiazTrigo2016, Parra2024}), and P-Cygni profiles appear in their optical and infrared emission lines (e.g. \citealt{MunozDarias2019}, \citealp{SanchezSierras2023}), indicating the presence of accretion disc winds (see \citealt{MunozDarias2026} for a review). These are thought to play a key role in the evolution of the outburst and, more generally, in the accretion process, as they can extract a significant amount of mass and momentum from the system (e.g. \citealt{Fender2016}, \citealt{Tetarenko2016}). 

In this context, the study of the 2015 outburst of the BHXB V404~Cygni (V404~Cyg hereafter) marked a significant breakthrough. This system, located at a distance of $d = 2.39\pm 0.14$\,kpc (\citealt{Miller2009}), hosts an $\sim9$\,M$_{\odot}$ black hole (\citealt{Khargharia2010}) in an orbital period of $\sim6.5\,$d (\citealt{Casares1992}).
V404~Cyg was first discovered during an outburst episode in 1989 (\citealt{Makino1989}). After 25 years of quiescence, the system entered a new outburst phase in 2015 (\citealt{Barthelmy2015, Kuulkers2015}). 

Extreme multi-wavelength variability was observed during the outburst (e.g. \citealt{Rodriguez2015}), including phases of enhanced X-ray obscuration (e.g. \citealt{Motta2017a, Motta2017b}). Conspicuous optical P-Cygni profiles were observed in at least twelve optical transitions, indicating outflow velocities of up to $\sim 3000$~km~s$^{-1}$ (\citealt{MunozDarias2016, MataSanchez2018}, hereafter referred to as \MD\ and \MS, respectively), consistent with those derived from contemporaneous X-ray observations, which also exhibit strong P-Cygni profiles (\citealt{King2015, MunozDarias2022}). These accretion disc wind signatures were primarily detected during the first $\sim 6$ days of the outburst, prior to the X-ray peak, when a likely super-Eddington plateau was observed (\citealt{SanchezFernandez2017}). From this point, the luminosity suddenly dropped to near-quiescence levels across all bands (e.g. \citealt{Sivakoff2015, Jourdain2017}), only $\sim 12$ days after the beginning of the outburst. This abrupt decay corresponds to the onset of the so-called nebular phase.

The nebular phase of V404~Cygni represents a unique observational case, characterised by extreme properties: optical recombination lines of an intensity unparalleled in similar systems, as evidenced by their exceptionally large equivalent widths (EWs; e.g. EW$_{\text{H}\alpha}\sim100-2000$\,\AA), Balmer decrements (BDs, defined as the H$\alpha$/H$\beta$ flux ratio) in the range of $\sim4-7$, and broad line wings reaching 3000~km~s$^{-1}$ (\MD, \citealt{Rahoui2017}, \MS). In addition, the flux ratio of \ion{He}{ii}-$4686$ to H$\beta$ (hereafter $I_{\mathrm{ratio}}$), which serves as a proxy for the ionisation state of the gas, is observed to decrease to values $\lesssim1$,  lower than those typically measured during outburst.

Taken together, these properties are naturally interpreted as follows. Once the luminosity drops, the ejecta expand and cool, providing the conditions for efficient recombination as they evolve from an optically thick to an optically thinner regime. This scenario is further supported by the agreement between the velocities inferred from the widths at the base of the lines during the nebular phase and those derived from the blue edge of the earlier P-Cygni profiles (e.g. \MD).
The nebular emission was detected for ten days and dominated the optical spectrum for about four days before gradually fading, while the accretion disc emission progressively re-emerged (\citealt{Casares2019}).

The study of the nebular phase enabled the estimation of the mass ejected by the wind. Using two independent approaches based on the determination of the diffusion and recombination timescales,  \MD\ and \cite{Casares2019} estimated consistent values for the outflow mass (M$_{\rm{out}}\sim10\elev{-5}-10\elev{-8}$ M$_{\odot}$ and  M$_{\rm{out}}\simeq4\times10\elev{-6}$ M$_{\odot}$, respectively). In both cases, M$_{\rm{out}}$ is of the order of or greater than the mass accreted during the outburst (\citealp{Zycki1999}; see also \MD), supporting the scenario in which winds are responsible for a significant depletion of the accretion disc and the sudden turn-off of the 2015 outburst.

Although strong emission line spectra such as those observed during the nebular phase of V404~Cyg are unique, evidence of a similar phenomenology has been detected in other X-ray binaries. A remarkable example is V4641 Sagittarii, whose spectra exhibit EW$_{\text{H}\alpha}\sim100$ Å and broad line wings extending up to 3000 km s$^{-1}$ (e.g.  \citealt{Chaty2003, Lindstrom2005, MunozDarias2018}). Nebular phases have been detected in at least two outbursts of this source, typically emerging after a highly luminous stage of the outburst and following the observation of P-Cygni line profiles. These profiles indicate the presence of optically thick winds (\citealt{MunozDarias2018}). These findings strengthen the hypothesis that the nebular phase corresponds to an optically thinner stage of massive disc winds.  The detection of the nebular phase in V404~Cyg and V4641 Sagittarii, along with similar — though weaker — spectral features observed in other systems (see e.g. \citealt{PanizoEspinar2021, MataSanchez2024}), suggests that optically thin expanding material may be a common phenomenon in disc-fed X-ray binaries.

Photoionisation simulations can reproduce the spectral properties of such ejecta. Once the wind parameters and irradiation conditions are specified, these models enable predictions of the emergent spectrum. In this work, we employed version 23.01 of the photoionisation code \textsc{Cloudy} (\citealt{Chatzikos2023}) together with the \textsc{PyCloudy} library (\citealt{Morisset2013}) to perform the first modelling of the nebular phase. Our aim was to reproduce the spectral features that characterise the nebular spectra of V404~Cygni and to investigate the physical conditions that enable this phase to persist over a timescale of several days. Finally, we used different approaches to estimate the mass ejected during the 2015 outburst.

\section{Observational data and methods} 
In this section, we present the observational data we aim to reproduce and the photoionisation models we employed for this purpose.

\subsection{Observed spectra}

\begin{table}
\caption{Daily ranges of BD, $I$\textsubscript{ratio}, and H$\alpha$ EW (\AA) during the nebular phase of V404~Cygni.}            
    \label{table_obs}      
    \centerline{        
    \begin{tabular}{c c c c}    
    \hline                           
     MJD (Day) & BD & $I$\textsubscript{ratio} & H$\alpha$ EW (\AA) \\ 
    \hline  
      57200 (10) & $>2.5$\tablefootmark{*} & $1.0-1.1$ & $>205$\tablefootmark{*} \\
      57200.27 (10)  &  $4.61$ & -- & $1129.0$ \\
      57201 (11) & $4.8-5.9$ & $0.06-0.12$ & $1472-1904$  \\
      57202  (12) &  $5.5-7.4$ & $0.03-0.3$ & $408-685$ \\ 
      57203  (13) &  $5.3$ & $0.2$ & $183-189$ \\
    \hline                  
    \end{tabular}}
    \tablefoot{
    Values are taken from the supplementary material (Table A5) of \MS, except for the second entry (MJD 57200.27), which corresponds to \citealt{Rahoui2017}.\\
    \tablefoottext{*}{Lower limits on the BD and H$\alpha$ EW are not included in this work.}
    }
\end{table}

The 2015 nebular phase of V404~Cyg began on June 27 and persisted for about 10 days. Here, we focus on the properties of the optical spectra during the interval in which they were dominated by nebular emission, namely from June 27 to June 30 (\citealt{Casares2019}; days 10--13 from the beginning of the outburst, hereafter referred to in this format). Most of the data referred to in this paper were collected with the Optical System for Imaging and low-intermediate Resolution Integrated Spectroscopy (OSIRIS; \citealt{Cepa2003}) instrument on the Gran Telescopio Canarias (GTC). In particular, out of the 651 spectra collected by GTC during the entire outburst, 62 were obtained during the nebular phase. 
These data were already presented in \MD\ and \MS, to which we refer the reader for a detailed description of the data reduction procedure and analysis. In particular, full details of the identified spectral features, as well as the daily ranges of the line flux ratios and EWs used in this work, are reported in \MS\ and its supplementary material.

\begin{figure}
\centering
\includegraphics[width=0.5\textwidth]{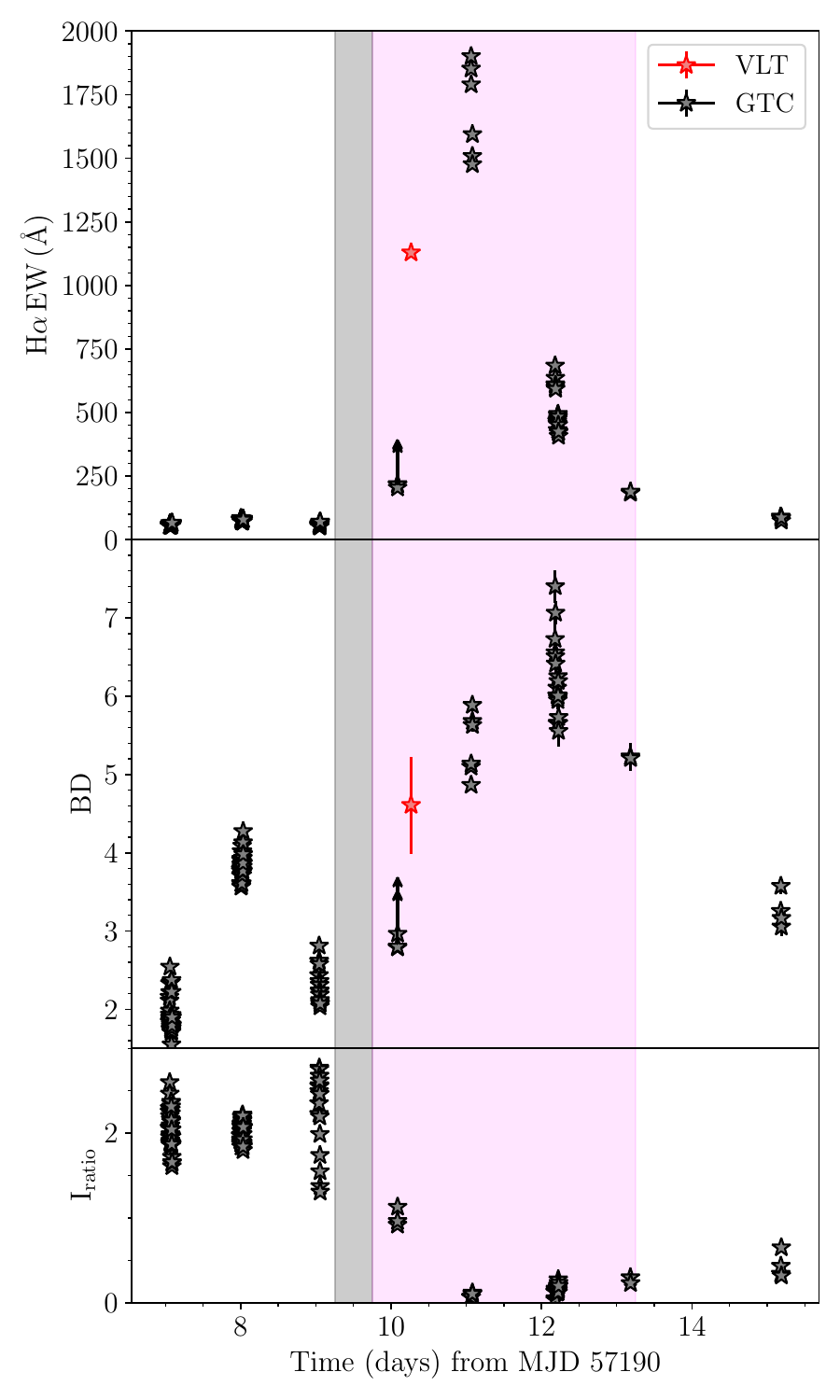}
\caption{Temporal evolution of the three line diagnostics used in this work (days 7 to 15 from the onset of the outburst). From top to bottom: H$\alpha$ EW, BD, and $I_{\mathrm{ratio}}$. The black symbols correspond to GTC data (\protect\MD, \protect\MS). The blue symbols indicate VLT values (\citealt{Rahoui2017}). The upward-pointing arrows denote lower limits. The grey-shaded region marks the plateau at the outburst peak preceding the rapid decay, which is followed by the nebular phase (pink-shaded region). The error bars are, in most cases, smaller than the symbols.}
\label{Fig:ParamsEvolution}
\end{figure}

Daily intervals of the relevant line diagnostics used in this work are provided in Table~\ref{table_obs}. Figure ~\ref{Fig:ParamsEvolution} shows the evolution of the H$\alpha$ EW, BD, and $I_{\mathrm{ratio}}$ during the three days preceding the nebular phase, as well as throughout it. The figure highlights an increase of almost two orders of magnitude in the EW, a rise in the BD to values $\gtrsim 4$, and a decrease in $I_{\mathrm{ratio}}$, immediately following the X-ray drop that marks the end of the X-ray plateau. We note that on day 10, the intensity of the Balmer emission lines saturated the detector. As a consequence, our data provide only lower limits for the H$\alpha$ flux and EW. These values were therefore discarded from our analysis. In addition to the GTC dataset, we include additional values reported by \cite{Rahoui2017}. These were derived from an optical spectrum obtained using the FORS2 medium-resolution spectrograph at the Very Large Telescope (VLT) on day 10, approximately 4.34 hours after the first GTC observation on the same day (see Table \ref{table_obs}).

\subsection{Photoionisation models}
\textsc{Cloudy} is a 1D photoionisation code designed to predict the spectrum emerging from a wide range of astrophysical plasmas (\citealt{Ferland1998}). By self-consistently solving the equations for statistical equilibrium, thermal equilibrium, ionisation-recombination and heating-cooling processes, the code calculates the thermal, ionisation and chemical structure of a plasma cloud. Given its ability to model recombination-dominated environments, \textsc{Cloudy} represents an ideal code for the study of the V404~Cyg nebular phase. 

To operate the code the user needs to specify the cloud’s geometry, composition and irradiation conditions. For our simulations we adopted a plane-parallel geometry and solar abundances (\citealt{Asplund2009}). For simplicity, we assumed the incident radiation field to be represented by a piecewise power law (see below). The distance between the gas slab and the irradiating source was held constant throughout the simulations. Initially, the gas density was assumed to be constant across the cloud, while clumpiness was taken into account in a subsequent step by defining a covering factor (see Sect. \ref{sec:dis:npmass}). The code ends the calculations when a user-defined stopping criterion is met. In this case, we stopped the simulations when the fraction of free electrons dropped below 2\%, indicating a region in which nearly all the gas has recombined.

 Although the physical scenario is that of expanding and cooling wind, we assumed that the timescale to reach the photoionisation equilibrium is shorter than the expansion timescale. This assumption allowed us to consider a plasma cloud in hydrostatic equilibrium. To capture the variations of its physical properties due to both the outburst evolution and the nebula expansion, we produced a grid of models sampling a wide range for three parameters: the spectral index of the incident radiation field ($\alpha$), the hydrogen density ($n_{\rm{H}}$) and the ionisation parameter ($\xi$).

\subsubsection{The incident radiation field}
The incident continuum is represented as a power law to reflect the approximately flat continuum observed in nebular-phase spectra (\MS). In addition, the best-fitting photon indices obtained from Swift observations of V404~Cyg collected a few hours before the first nebular-phase spectrum analysed in this work and after the last flare of the outburst, lie in the range $\Gamma\approx1-2$ (\citealt{Motta2017a}). This corresponds to spectral indices in the range $\alpha\approx-1$ to 0.
We therefore sampled a grid of spectral index values around the flat-slope continuum: $\alpha = [-1,\, -0.5,\, 0,\, 0.5,\, 1]$. 

The adopted incident radiation field is a piecewise power-law function provided by the \textsc{table power law} command in \textsc{Cloudy}. It is defined as
\[\begin{cases} 
    f_{\nu}\propto \nu\elev{5/2}\text{ if }\nu< 3\times 10\elev{13}\,\text{Hz}\\
    f_{\nu}\propto \nu\elev{\alpha} \text{ if }3\times 10\elev{13}\,\text{Hz}\leq \nu\leq1.2\times10\elev{19}\,\text{Hz}\\
    f_{\nu}\propto\nu\elev{-2} \text{ if }\nu>1.2\times10\elev{19}\,\text{Hz}
\end{cases} 
\]
This definition prevents stability issues that may appear during the simulations. The cutoff at lower frequencies ($< 3\times 10\elev{13}\,\text{Hz}$, $\sim$10\,$\mu$m) limits the amount of free-free heating, while the slope at the highest frequencies ($>1.2\times10\elev{19}\text{Hz}$, $>50$\,keV) avoids a non-physical high-energy continuum.

\subsubsection{Density}\label{density}
Visual inspection of the spectra reported by \MS\ shows that weak forbidden lines, such as [\ion{O}{iii}]-4959, 5007 and [\ion{S}{ii}]-6717, 6731, are present during the nebular phase of V404~Cyg.  However, the properties of these lines are consistent with an origin in a low-density external shell \citep{Casares2019}. The absence of potentially stronger forbidden lines such as [\ion{O}{i}]-6300 and [\ion{O}{iii}]-4363 suggests that, for the ionisation range considered in our models (see below), the electron density of the emitting regions exceeds the critical value $n_c\sim10\elev{8}\,$cm$\elev{-3}$ (\citealt{Osterbrock2006}). We used this value as a lower limit for the hydrogen number density of our models.

As the nebular phase is thought to originate from optically thin ejecta, we considered the typical density of optically thick accretion disc winds in X-ray binaries as the higher limit for the gas density. This value has been estimated for a few sources through line ratios diagnostics in the X-ray band, yielding $n\sim10\elev{14-15}\,$cm$\elev{-3}$  (e.g. \citealt{Schulz2008, Kallman2009}). We therefore explored hydrogen density values in the range $n_{\rm{H}}=10\elev{8}-10\elev{14}\,$cm$\elev{-3}$, with a step of 0.5 dex.

\subsubsection{Ionisation parameter}\label{ionisation}
The ionisation parameter is defined as $\xi= L/(n \,r\elev2)$ and is expressed here in units of \(\mathrm{erg\,cm\,s^{-1}}\), which is omitted hereafter. Here, $L$ is the ionising luminosity and $r$ is the distance between the radiation source and the plasma cloud (\citealt{Tarter1969}). The ionisation parameter is tightly connected to the presence of ionised atomic species in the emitting regions. The observation of \ion{H}{i}, \ion{He}{i} and \ion{He}{ii} emission lines in the nebular phase spectra implies that hydrogen and helium cannot be heavily over-ionised. We therefore set an upper limit of $\log\xi =2$ to the ionisation parameter (\citealt{Kallman1982}).

Conversely, the presence of the Bowen blend (consisting of blended emission features from \ion{N}{iii}, \ion{C}{iii}, \ion{C}{iv} and \ion{O}{ii}; e.g. \citealt{Schmidtobreick2003}), observed in some nebular phase spectra, suggests $\log\xi\geq-1$. We used a conservative grid for the ionisation parameter, spanning from $\log\xi = -2$ to $\log\xi = 2$, with a step of 0.5 dex.

\begin{figure*}
\centering
\includegraphics[width=\textwidth]{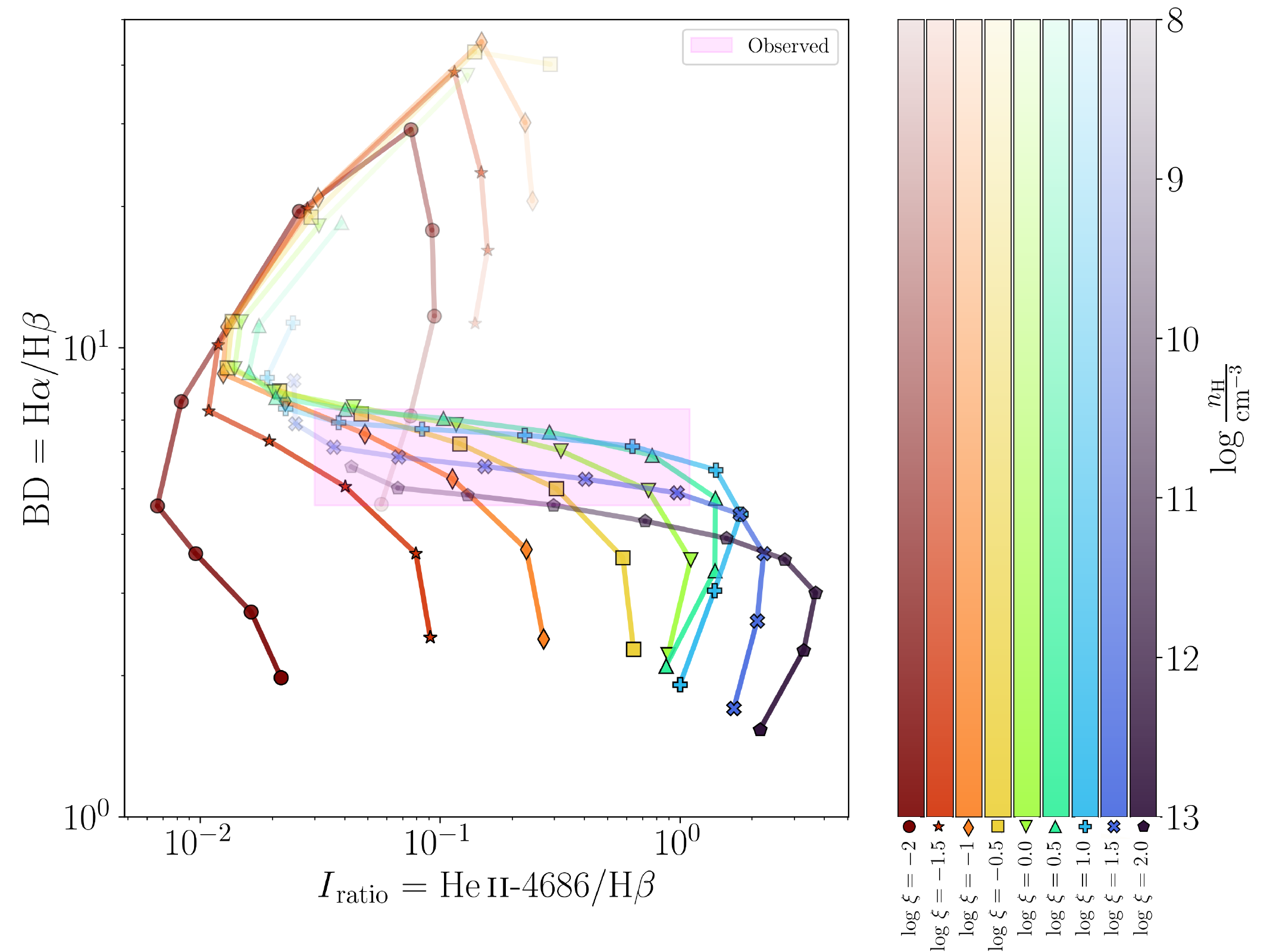}
\caption{BD vs $I_{\mathrm{ratio}}$ for the grid of models with a spectral index $\alpha = -0.5$. Models with the same ionisation parameter ($\xi$) are plotted with identical symbols and colours and connected by lines. The colour scale traces the ionisation parameter: the redder symbols correspond to models with lower ionisation parameters, while the bluer symbols indicate higher ionisation models. The opacity of symbols and lines increases with density. Models with densities $n_{\rm{H}} > 10^{13}$~cm$^{-3}$ are omitted, as their BDs are significantly lower than the observed values. The pink-shaded region marks the range of the observed values during the nebular phase. Several models with densities in the range $10^{8.5}$~cm$^{-3} \leq n_{\rm{H}} \leq 10^{12}$~cm$^{-3}$ and ionisation parameters $\log\xi \geq -1.5$ lie on a \textit{horizontal branch}, clustering within the observed region.}
\label{Fig:Iratio}
\end{figure*}

\section{Results}

The simulation yields 568 successfully calculated models out of 585. Most of the aborted simulations (13 out of 17) are characterised by low densities ($n \leq 10\elev9 {\text{cm}^{-3}}$) and high ionisation parameters ($\log\xi \geq 1$), conditions under which the ionisation structure of the cloud did not converge within a reasonable number of iterations. Among the remaining aborted simulations, all have $\alpha \geq 0.5$ and were terminated due to strong maser action. 

For every model in the grid, we derived the theoretical values of three key observables:
\begin{itemize}
    \item the BD, which provides insights into the dominating line emission processes and the density of the emitting regions (e.g. \citealt{Drake1980});
    \item the $I_{\rm{ratio}}$,  which provides information on the ionisation state of the gas (e.g. \citealt{Groot2001});

    \item the EW of the H$\alpha$ line, that is, the relative strength of H$\alpha$ with respect to the continuum.
\end{itemize}
These quantities were then compared with the observed values (i.e. Table \ref{table_obs}).

\subsection{The BD and the $I$\textsubscript{ratio}}
\label{sec:BD_Ir}
Figure \ref{Fig:Iratio} shows the predicted BD and $I$\textsubscript{ratio} as a function of density for the subset of models with spectral index of $\alpha = -0.5$. This value corresponds to the midpoint of spectral indices capable of reproducing the observations. 

In the upper region of the diagram, models with low ionisation parameters ($\log\xi \lesssim 0$) exhibit BDs that increase with density, reaching extreme values of several tens, with the density encoded in colour intensity along each ionisation track. Such high BDs are occasionally observed in accreting systems (e.g. novae and symbiotic binaries; \citealt{Mikolajewska1999, Mason2010}) and are commonly attributed to Balmer self-absorption and collisional excitation (e.g. \citealt{Netzer1975, Drake1980}), both of which enhance H$\alpha$ relative to H$\beta$. In this regime, the $I$\textsubscript{ratio} generally decreases with increasing density: while both \ion{He}{ii}-4686 and H$\beta$ fluxes rise with $n_{\rm{H}}$, H$\beta$ increases more rapidly. This is expected at low ionisation, where the neutral hydrogen fraction responsible for H$\beta$ emission is substantially higher than the fraction of singly ionised helium producing \ion{He}{ii}-4686.

Below a turning point at BD $\approx 9$ and $I_{\mathrm{ratio}} \sim 10^{-2}$, the models define a nearly horizontal region, hereafter referred to as the \textit{horizontal branch}. This region encompasses most intermediate-density models ($n_{\rm{H}} \sim 10^{10}$–$10^{12}$~cm$^{-3}$). Along this plateau, nearly constant BD values of $\sim 4$–$7$ are found across a range of models spanning different densities and ionisation parameters. These values exceed those predicted by Case B recombination, indicating additional radiative processes at play. The enhancement is consistent with Balmer self-absorption and collisional excitation, mechanisms often invoked to explain similarly high BDs in other accreting systems (e.g. \citealt{Ferland1978, Iijima2003}). The $I$\textsubscript{ratio} spans approximately two orders of magnitude and increases with the ionisation parameter, as higher ionisation favours a larger fraction of He \textsc{ii}, thereby enhancing He \textsc{ii}-4686 recombination emission relative to H$\beta$.

In the bottom part of the figure, high-density ($n_{\rm{H}}\sim10\elev{13}~{\text{cm}^{-3}}$) models show BDs in the range of $\sim 1$–$2$. Such flat decrements are expected when collisional de-excitation dominates over radiative transitions (e.g. \citealt{Adams1974}). Similar conditions have been proposed to explain the flat BDs occasionally observed in novae,  cataclysmic variables (e.g. \citealt{Kiplinger1979, Williams1980}) and tidal disruption events (e.g. \citealt{Short2020}). In this regime, the BD becomes nearly independent of $\xi$, while $I$\textsubscript{ratio} continues to increase with the ionisation parameter.

Consistent with the colour coding adopted in Fig.~\ref{Fig:ParamsEvolution}, the pink rectangle in Fig.~\ref{Fig:Iratio} delimits the parameter space covered by observations during the nebular phase (days 10 to 13). Within this period, the ranges for BD and $I_{\rm{ratio}}$ are $4.6-7.4$ and $0.03-1.1$, respectively. A wide range of models reproduce these observables. Most of these models correspond to densities in the range $10\elev{8.5}$\,cm$\elev{-3}\leq n_{\rm{H}}\leq10\elev{12}$\,cm$\elev{-3}$ and lie on the horizontal branch. 

The results for all spectral indices explored in the grid (i.e. $\alpha=-1$ to $\alpha=1$) are presented in  Fig.~\ref{BD_Iratio_big} in the appendix. The models exhibit a distribution in the BD–$I$\textsubscript{ratio} plane similar to that discussed previously for the representative case with a spectral index $\alpha = -0.5$. Most models with densities in the range $n_{\rm{H}} \sim 10^{10}$–$10^{12}$ cm$^{-3}$ lie along a horizontal branch, characterised by approximately constant BD values while the $I$\textsubscript{ratio} spans several orders of magnitude. High-density models ($n_{\rm{H}} \sim 10^{13}$ cm$^{-3}$) occupy the lower region of the plots, where the BD decreases as collisional processes become dominant. Among the low-ionisation models ($\log\xi < 0$), several exhibit extreme BD values (BD $\gtrsim 10$), consistent with those seen in Fig.~\ref{Fig:Iratio} for $\alpha=-0.5$.

Figure~\ref{BD_Iratio_big} also shows that increasing the spectral index results in a broader plateau, with lower $I$\textsubscript{ratio} values reached as $\alpha$ increases. In the simulations with spectral indices in the range $-1 \leq \alpha \leq 0$, several models along the horizontal branch reproduce the observed BD and $I$\textsubscript{ratio} values. Conversely, models with $\alpha \geq 0.5$ are inconsistent with the observations.

\subsection{The \texorpdfstring{H$\alpha$}{Ha} equivalent width}

\begin{figure*}
\centering
\includegraphics[width=\textwidth]{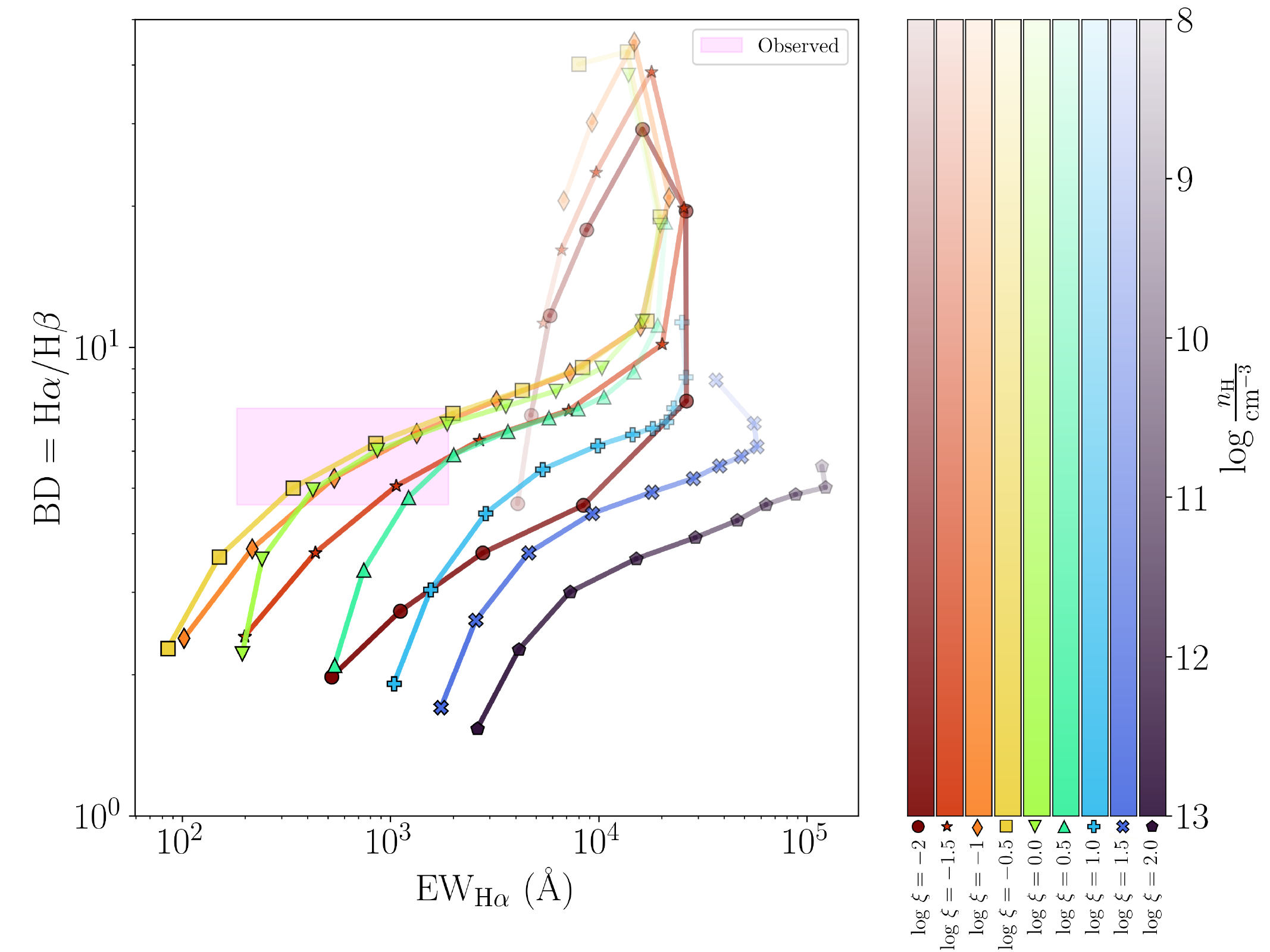}
\caption{BD vs H$\alpha$ EW for the grid of models with fixed spectral index $\alpha=-0.5$. Symbols and colours as per Fig.~\ref{Fig:Iratio}. A subset of models, characterised by intermediate densities  $ 10^{11}$ cm$^{-3}\leq n_{\rm{H}}\leq 10^{12}$ cm$^{-3} $ and ionisation parameters $-1.5\leq\log\xi\leq0.5$, reproduce the observed values.}
\label{Fig:EW}
\end{figure*}

For each model, we computed the predicted EW of the H$\alpha$ line following the \textsc{Cloudy} prescription\footnote{ \textsc{Cloudy} manual Hazy 2, §2.3}:
\begin{equation}
EW_{\rm{H}\alpha} = \lambda \frac{F_{\rm{H}\alpha}}{\lambda F_\lambda\elev{\rm{cont}}},
\end{equation}
where we omitted the conventional negative sign to enable a direct comparison with observations. In this expression, 
$\lambda=6563$\AA\, is the H$\alpha$ wavelength, $F_{\rm{H}\alpha}$ is the line flux, and $\lambda F_\lambda\elev{\rm{cont}}$ is the transmitted continuum flux at the same wavelength, as provided by the code. The results for the entire model grid, plotted as a function of the BD, are shown in Fig.~\ref{BD_EW_big} in the appendix. Figure~\ref{Fig:EW} presents the subset of models computed with a spectral index $\alpha = -0.5$. The predicted H$\alpha$ EWs span several orders of magnitude, from $\sim10^{2}$\,\AA\, to $\sim10^{5}$\,\AA.

In the upper region of the diagram, corresponding to large BD values, models with low ionisation parameters ($\log\xi \lesssim 0$) show increasing EW with increasing density. This behaviour can be attributed to Balmer self-absorption and collisional excitation, which enhance the emergent H$\alpha$ emission relative to both the continuum and H$\beta$. The EW rises with density up to $\sim10^{4}$\,\AA, reaching $\sim10^{5}$\,\AA\, in the most highly ionised models. 

In the central and lower regions of the diagram,  where most models cluster, the trend reverses: the EW decreases with increasing density. This inversion likely reflects both the enhancement of the recombination continuum and the increasing importance of collisional de-excitation in the highest density models.

Most models characterised by intermediate densities ($n_{\rm{H}}\sim 10^{10}-10^{12}~{\rm cm}^{-3}$) occupy the central region of the diagram, where BD values of $\sim4-10$ and EWs from a few hundred to $10^{5}$\,\AA, are predicted. The BD and H$\alpha$ EW observed during the nebular phase of V404~Cyg fall at the low-EW end of this region. Despite variations of several orders of magnitude in EW, the restricted range of BD values indicates that the line-emitting gas maintains relatively stable excitation conditions over a wide range of density, ionisation, and irradiation, while the absolute line and continuum fluxes evolve significantly. Models with densities of $n_{\rm{H}}\sim10^{11}-10^{12}\,{\rm cm}^{-3}$ and ionisation parameters $-1.5\lesssim\log\xi\lesssim0.5$ reproduce the observed values (pink shaded region in Fig. \ref{Fig:EW}).
\smallskip

\begin{figure*}
\centering
\includegraphics[width=\textwidth]{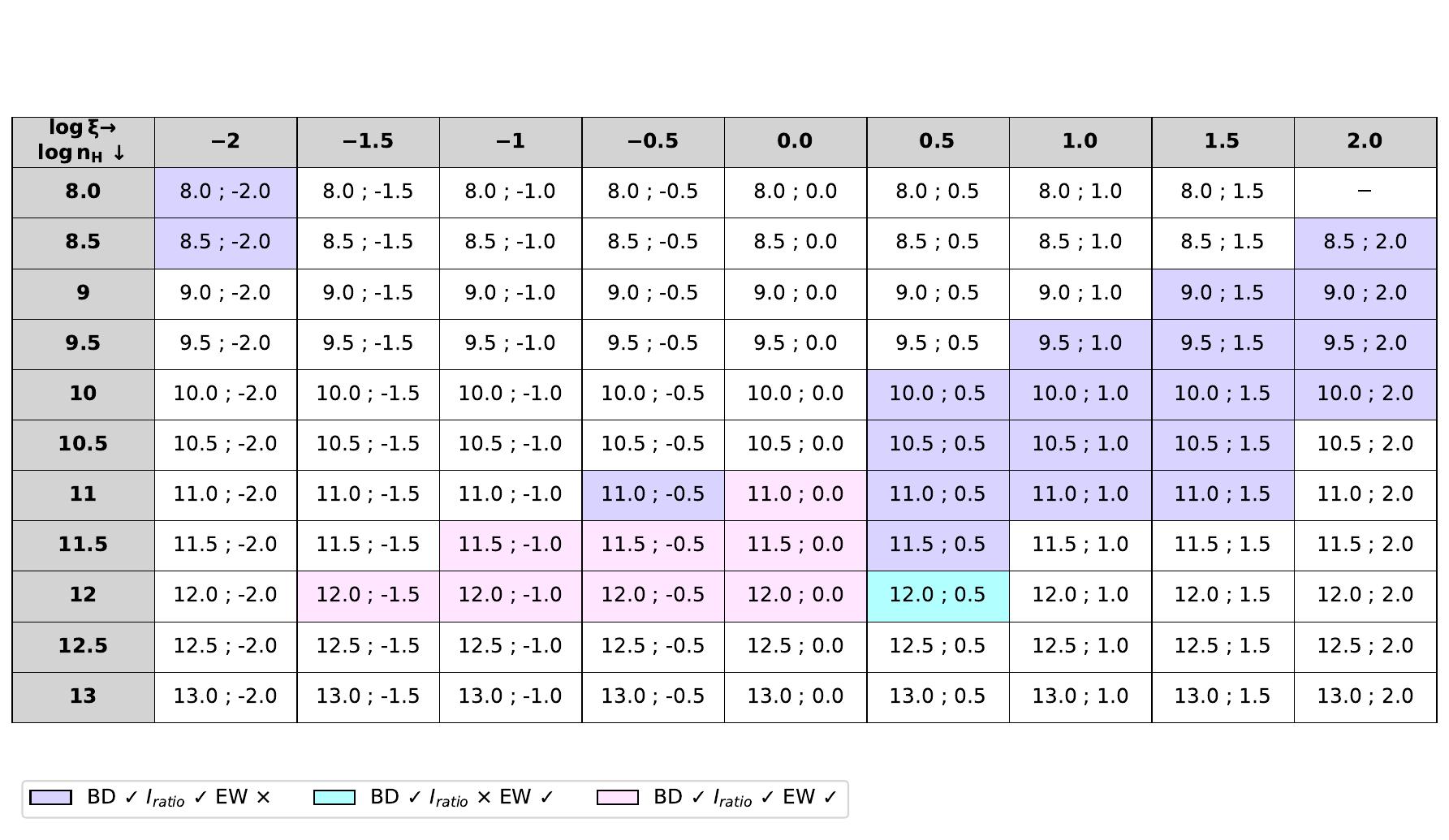}
\caption{Model grid for the spectral index $\alpha = -0.5$. The models highlighted in violet simultaneously reproduce the observed values of BD and $I_{\mathrm{ratio}}$. Those in cyan reproduce BD and EW$_{\mathrm{H\alpha}}$ and those in pink reproduce all three quantities simultaneously. Models labelled with an em dash `--' indicate aborted simulations.}
\label{BDIratioEW_chart-0.5}
\end{figure*}

Figure~\ref{BDIratioEW_chart-0.5} lists the models that simultaneously reproduce the observed BD, $I_{\mathrm{ratio}}$, and H$\alpha$ EW for the subgrid with a spectral index $\alpha = -0.5$. These observables are matched by a set of models with intermediate densities ($10^{11}\,{\rm cm}^{-3} \lesssim n_{\rm{H}} \lesssim 10^{12}\,{\rm cm}^{-3}$) and ionisation parameters $-1.5 \lesssim \log\xi \lesssim 0$, most of which lie on the horizontal branch in the BD--$I_{\mathrm{ratio}}$ plane (see Sec.~\ref{sec:BD_Ir} and Fig.~\ref{Fig:Iratio}).

Simulations with other spectral indices display similar qualitative trends (Appendix~\ref{appendice_BDEW}). In all cases, the H$\alpha$ EW exhibits a strong dependence on density, decreasing for the densest models. Each subset includes models that lie within the observed region. However, only simulations with spectral indices $\alpha \leq 0$ yield models simultaneously consistent with the observed BD, $I$\textsubscript{ratio} and EW  values (see the charts in Appendix~\ref{appendix_table}).

\begin{figure*}
\centering
\includegraphics[width=\textwidth]{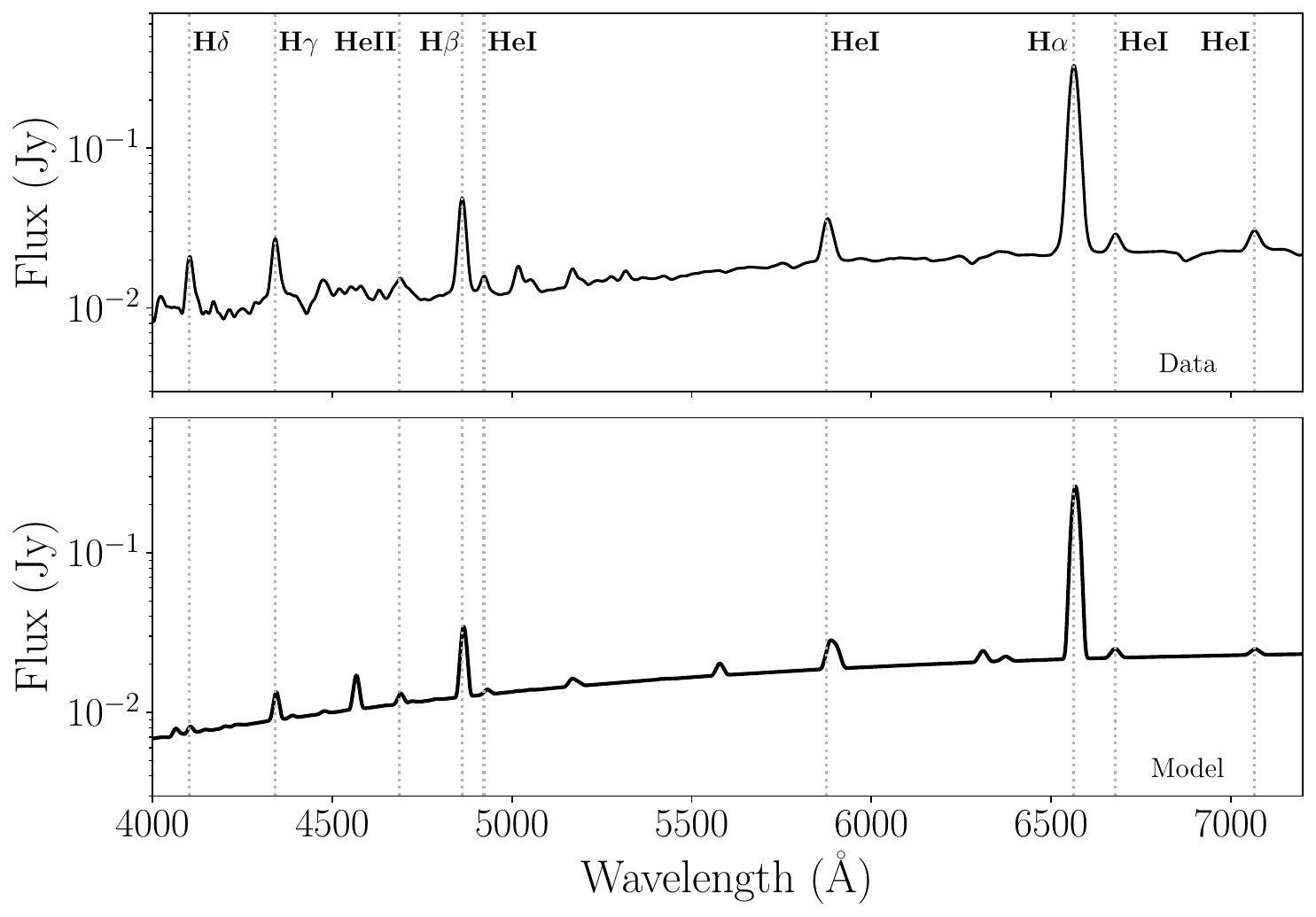}
\caption{Top panel: V404~Cyg optical spectrum obtained at MJD~57202.214 (day 12), smoothed to match the Cloudy resolution. The most prominent emission lines are marked. Bottom panel: optical spectrum obtained from modelling the nebular phase with \textsc{Cloudy}, assuming a spectral index $\alpha = -0.5$, $\log\xi = 0$ and $\log n_{\rm{H}} = 11$. The model spectrum was normalised to match the observed continuum flux density adjacent to the H$\alpha$ line. This exemplifies how a simple photoionisation model can broadly reproduce the main spectral properties of the nebular phase. Overall, the synthetic spectrum matches the continuum slope and reproduces the main hydrogen and helium lines. Larger differences appear for H$\gamma$ and H$\delta$, where the S/N of the observed spectrum is lower. In both spectra, the flux is shown on a logarithmic scale owing to the intensity of the emission lines, particularly H$\alpha$.}
\label{spectrum}
\end{figure*}

\section{Discussion}\label{discussion}

In this study, we modelled the nebular phase observed at the end of the 2015 outburst of V404~Cyg using the widely tested photoionisation code \textsc{Cloudy}. The aim of this work was to constrain the range of physical conditions in the expanding gas which allow this phase to dominate the optical spectrum of the BHXB over several days. We explored an extensive grid of photoionisation models spanning a wide range of gas densities, ionisation parameters, and spectral indices of the illuminating radiation field. A subset of these models reproduces the key observables. Below, we discuss the physical implications of our results and the limitations of our modelling, as well as provide estimates of the outflow mass responsible for the nebular phase and its structure.

\subsection{Physical parameters of the emitting gas}
\label{subsec:physical}
The models that simultaneously reproduce the observed values of BD, $I$\textsubscript{ratio}, and H$\alpha$ EW  correspond to spectral indices $\alpha \leq 0$ (see Appendix~\ref{appendix_table}). These models span a broad range of densities and ionisation parameters, indicating that the nebular phase observed in V404~Cyg can arise and persist under diverse, physically plausible conditions. 

Relatively stable trends can be identified in the BD–$I_{\mathrm{ratio}}$ and BD–EW diagrams, most clearly in the so-called horizontal branch in the former case. Observed values lie in these regions, where models  cluster (see Fig. \ref{Fig:Iratio} and \ref{Fig:EW}). This qualitatively explains why the observables of the nebular phase evolve smoothly over day-long timescales despite orders-of-magnitude variations in the density and ionisation parameter. Such variations are expected as the ejecta expand, cool and recombine. They are naturally reproduced by our modelling, which shows that they can translate into a relatively narrow range of observable properties. Given that wind-type outflows are a common phenomenon in BHXBs (e.g. \citealt{MunozDarias2026}), our results suggest that the nebular phase may represent a common stage in the post-outburst evolution of these systems. 
\smallskip

A more quantitative interpretation can be obtained by selecting the subset of models that simultaneously reproduce all three diagnostic observables. This yields the following physical constraints for the outflow.

\begin{itemize}
    \item \textbf{Density:} The models that reproduce the observables lie within the range $10^{8}\,$cm$^{-3} \lesssim n_{\rm H} \lesssim 10^{12}$\,cm$^{-3}$.
Models near the lower edge of this interval ($n_{\rm H} \sim 10^{8}$–$10^{8.5}$\,cm$^{-3}$) can reproduce the observed flux ratios and EWs. However they do not lie along the horizontal branch identified in Fig.~\ref{Fig:Iratio} and are therefore inconsistent with the persistence of the nebular phase. We thus focus on densities within $10^{10.5} \lesssim n_{\rm H} \lesssim 10^{12}$\,cm$^{-3}$, which both match the observables and lie on the horizontal branch. These densities are somewhat lower than typically inferred for optical winds in BHXBs, where both observations and simulations suggest $n_{\rm H} \gtrsim 10^{13}$ cm$^{-3}$ for the line-emitting regions (\citealt{Rahoui2014, Koljonen2023}). However, they are consistent with the picture of an expanding, partially optically thin phase of an optical wind, likely representing the later evolution of the same ejecta observed at the outburst peak. In this phase, Balmer self-absorption and collisional excitation likely remain significant, while recombination dominates both the cooling and the emission. By contrast, during the earlier, optically thick stages of the wind, higher densities and stronger radiative transfer effects are expected to shape the optical spectra. The densities recovered here could therefore be characteristic of BHXBs nebular phases following the dense wind stage.

\item \textbf{Ionisation parameter:}  
The models that reproduce the observables lie in the range $-1.5 \lesssim \log\xi \lesssim 1$, typical of gas where hydrogen and helium are only partially ionised, as required to produce strong Balmer and He \textsc{i} lines (\citealt{Kallman1982}). Similar ionisation levels are inferred for the optical-emitting zones of MAXI~J1820+070 (\citealt{Koljonen2023}).
\item \textbf{Spectral index:}  
Only models with flat or negative spectral indices ($\alpha \leq 0$) reproduce the observations.  Although representing the incident spectrum as a single stationary power law is clearly a simplification (see below), the fact that models with different $\alpha$ values still reproduce the observables indicates that the emission diagnostics are not overly sensitive to moderate variations in the spectral slope. 
\end{itemize}

Adopting a representative model within the allowed region ($\alpha=-0.5$, $n_{\rm H}=10^{11}$ cm$^{-3}$, $\log\xi=0$) yields a synthetic spectrum that qualitatively matches that of V404~Cyg during the nebular phase [see Fig.~\ref{spectrum} for day 12 (MJD~57202.214)]. For direct comparison, the observed spectrum was smoothed to match Cloudy resolution (R = 300). The model spectrum was normalised to the observed continuum flux density adjacent to the H$\alpha$ line and convolved with a Gaussian profile with a full width at zero intensity of 1000 km s$^{-1}$ to mimic the observed spectral lines. Qualitatively, the model reproduces both the dominant emission lines and their relative strengths remarkably well. Despite the simplified treatment of the continuum, the overall spectral shape is also well matched, with the model showing a similar rise towards redder wavelengths.

\subsection{Limitations}\label{limitations}

Our modelling reproduces the key observables of the nebular phase. However, several limitations must be noted.

First, models at the edge of the explored parameter space (i.e. those with $\log\xi \geq 1$, $n_{\rm H} = 10^{8}$, or a combination of both) were formally marked as converged but exhibited behaviour consistent with the limits of \textsc{Cloudy}’s applicability.
At high $\xi$ values, the solutions became optically thick to electron scattering, while in cases combining high ionisation and low density, the stopping criterion, requiring 2\% of free electrons, was not reached within a reasonable number of zones. Results in this regime should therefore be interpreted with caution. However, they do not impact the conclusions drawn from the main parameter range explored. 

In addition, the use of a power law to describe the ionising continuum is likely the main source of uncertainty. The nebular phase occurred when the X-ray flux from V404~Cyg  abruptly dropped, initiating the transition towards near-quiescent levels (e.g. \MD), and variations in the adopted spectral energy distribution can significantly affect the model output. This introduces uncertainties in quantities sensitive to the continuum level, such as the H$\alpha$ EW, and should be kept in mind when interpreting these results. Nevertheless, the existence of stable solutions for a subset of spectral indices suggests that our conclusions are not strongly dependent on the detailed shape of the spectral energy distribution.

Finally, our use of a static cloud instead of  expanding wind introduces simplifications. This approximation neglects Doppler broadening, which redistributes line photons over wavelength but does not alter the total line flux or the flux ratios central to our analysis. In addition, the assumption of a plane-parallel geometry is expected to introduce only second-order effects in the resulting spectra (\citealt{Ferland2006}), which are not expected to alter our main conclusions.

\subsection{The mass involved in the nebular phase}
\label{sec:dis:npmass}

Estimating the outflow mass is fundamental for understanding the impact of winds on the accretion process, yet observational constraints remain scarce (see e.g. \citealt{MunozDarias2026}). There are several reasons for this. First, the wind launching mechanism is not yet fully understood, introducing significant uncertainty into mass-loss predictions, as they depend on the assumed driving force (e.g. \citealt{Dubus2019}). Second, in the case of optically thick winds, which are typically detected via P-Cyg line profiles, scattering and absorption processes require detailed radiative transfer modelling to accurately interpret observations. However, once the ejecta become optically thinner, the analysis can be simplified. This is the case for the 2015 nebular phase of V404~Cyg. In this regime, the optical emission-line spectrum is expected to be dominated by recombination, self-absorption, and collisional effects, which can be modelled with photoionisation codes such as \textsc{Cloudy}.

In this context, order-of-magnitude estimates informed by the physical properties derived in this work can still provide valuable insights. As a first step, we estimated the mass involved in the nebular phase using two different approaches.

\subsubsection{Volume-dependent mass estimate}\label{volumebased_estimate}
We first derived a simple constraint using a volume-based approach. We assumed that, during its expansion, the ejecta occupy an overall volume with an average gas density constrained by our photoionisation models. The nebular phase is characterised by strong emission lines, with broad wings extending up to $\pm3000$~km~s$^{-1}$. As discussed in Sect. \ref{sec:intro}, this velocity is consistent with the wind velocity inferred from the P-Cygni line profiles detected earlier in the outburst. We adopted this value as representative of the characteristic outflow expansion velocity ($v_{\rm out}$). Taking the $e$-folding timescale of the nebular emission, $t=0.7$~d (\citealt{Casares2019}), as representative of the timescale over which the ejecta remain observable, we estimate a characteristic radius of $R = v_{\rm out}\times t\simeq2\times10^{13}$\,cm. The total mass contained within this volume can be then approximated by
\begin{equation}\label{mass_out}
M_{\rm vol} = \rho\,  V = \rho f_{\rm C}\,\frac{4}{3}\,\pi \,R^3,
\end{equation}
where $\rho = n_{\rm H}\,m_{\rm H} + n_{\rm He}\,m_{\rm He} = 1.4\,n_{\rm H}\,m_{\rm H}$ represents the average gas mass density within the expanded volume (assuming solar abundances) and $\frac{4}{3}\,\pi \,R^3$ is the volume\footnote{The volume approximation used in Eq.~\ref{mass_out} assumes that the wind is launched at a radius $R_{\rm l}$ close to the compact object. Previous observational studies constrain the launch radius to $R_{\rm l}\sim10^{10}$\,cm (\citealt{MunozDarias2016}). Since $R_{\rm l}\ll R$, we neglected the launch radius when estimating the outflow volume. } 
of a sphere of radius $R$. In addition, motivated by the observational evidence for a preferentially equatorial outflow geometry in BHXBs (e.g. \citealt{Ponti2012, DiazTrigo2016, MunozDarias2026}), we assumed that, rather than being purely spherical, the ejecta cover a fraction $f_{\rm C}$ of the solid angle. We adopted the simple case with a half-opening angle of $30^{\circ}$, which corresponds to $f_{\rm C} \simeq 0.5$. We note that this assumption has no significant effect on this order-of-magnitude estimate.

We obtain $M_{\rm vol}\sim 5\times10^{-7} - 1\times10^{-5}\,M_{\odot}$ for $n_{\rm H}$ in the range $10^{10.5}-10^{12}$\,cm$^{-3}$ (see Sect.~\ref{subsec:physical}). This estimate can be related to the mass of the ejecta, $M_{\rm out}$, by considering the filling factor of the emitting volume, $f_{\rm V}$, such that $M_{\rm out} = f_{\rm V} M_{\rm vol}$. As discussed below, increasing evidence indicates that BHXB disc winds are highly structured, with significant clumping and thus $f_{\rm V} < 1$.

\subsubsection{Ionised mass from recombination}\label{recomb_estimate}
A complementary, volume-independent estimate can be obtained from the H$\alpha$ recombination luminosity, which traces the ionised gas mass. This approach is commonly employed to constrain the mass of the broad-line region in active galactic nuclei \citep[e.g.][]{Baldwin2003, Abolmasov2017, Zhang2024}, where similar electron densities are typically inferred.

Following \citet[Ch.~13.5]{Osterbrock2006}, the H$\alpha$ recombination luminosity can be expressed in terms of the effective recombination coefficient $\alpha^\text{eff}_\mathrm{H\alpha}$ as
\begin{equation}\label{Mion1}
  L_\mathrm{H\alpha} = h \nu_\mathrm{H\alpha} \int_V n_\text{e} n_\text{p} \alpha^\text{eff}_\mathrm{H\alpha} dV \approx  h \nu_\mathrm{H\alpha} n_\text{e} n_\text{p} \alpha^\text{eff}_\mathrm{H\alpha} V, 
\end{equation}
where $n_\mathrm{e}$ and $n_\mathrm{p}$ are the free electron and ionised hydrogen number densities, respectively, $h\nu_\mathrm{H\alpha}$ is the energy of an H$\alpha$ photon, and the second equality assumes uniform density throughout the emitting volume $V$. We expressed Eq.~\ref{Mion1} in terms of the ionised mass $M_\mathrm{ion} = 1.4\,n_\text{p} m_\mathrm{p} V$, where we assumed solar abundances. We also rewrote the the H$\alpha$ luminosity in terms of its observed flux, $F_\mathrm{H\alpha} = 
L_\mathrm{H\alpha} / 4\pi d^2$, where $d = 2.39\pm 0.14$\,kpc is the distance to the source (\citealt{Miller2009}).  We obtain 

\begin{equation}\label{Mion2}   
  M_\text{ion} \sim 1.4 m_\text{p} \frac{4 \pi d^2 F_\mathrm{H\alpha}}{h \nu_\mathrm{H\alpha} n_\text{e} \alpha^\text{eff}_\mathrm{H\alpha}}.
\end{equation}

This estimate depends on the local electron density and temperature through 
$\alpha^\mathrm{eff}_\mathrm{H\alpha}$, but is independent of the spatial distribution of the emitting gas. While the ejecta is expected to contain both ionised and neutral material during the recombination process (see below), our photoionisation models indicate that the H$\alpha$-emitting gas is almost fully ionised. Thus, we can assume this mass is responsible for the emission over the nebular phase $e$-folding timescale of $t\approx0.7$~days.

We evaluated Eq.~\ref{Mion2} for each of our models, adopting the maximum H$\alpha$ flux observed during the nebular phase, $F_{\mathrm{H}\alpha}=4.3\times10^{-11}$\,erg\,cm$^{-2}$\,s$^{-1}$. The effective recombination coefficient $\alpha^\mathrm{eff}_\mathrm{H\alpha}$ was computed with \textsc{PyNeb} \citep{Luridiana2015} using the electron temperature $T_{\rm e}$ and free electron density $n_\text{e}$ evaluated at the location of peak H$\alpha$ emissivity. Where $T_{\rm e}$ exceeds the upper boundary of the \textsc{PyNeb} tabulation  ($3\times10^{4}$~K), we adopted this value as a conservative upper limit. This does not affect the models reproducing the observables, and therefore does not influence our conclusions.  The results for the full grid of models are presented in Appendix~\ref{appendice_mass}.

For models consistent with the observables, the free electron densities and temperatures span the ranges $n_\text{e}=10^{10.5}-10^{12.5}$\,cm$^{-3}$ and $T_{\rm e}=1-3\times10^4$\,K, yielding ionised masses $M_{\rm ion}\sim2 \times 10^{-11}-7 \times 10^{-9}\,M_{\odot}$. These values relate to the mass of the ejecta through the ionisation fraction, $f_{\rm ion} \leq 1$, which represents the fraction of ionised mass in the ejecta, yielding $M_{\rm out} = M_{\rm ion}/f_{\rm ion}$.

We note that the above estimate assumes that the H$\alpha$ emission arises entirely from case~B recombination. However, at the high densities ($\log n_\mathrm{e} \gtrsim 8 \text{ cm}^{-3}$) considered in our modelling, Balmer self-absorption and collisional excitation can contribute to the H$\alpha$ line emission \citep{Netzer1975, Drake1980}, potentially biasing this estimate. As an additional check, we therefore estimated the mass contributing to H$\alpha$ emission, adopting the H$\alpha$ emissivities computed by \textsc{Cloudy}, which accounts for these effects (see Appendix~\ref{App:mass_from_emissivity}). This approach yields  $M_{\rm emis}\sim$ $9 \times 10^{-12}-7 \times 10^{-9}\,M_{\odot}$, in agreement with the recombination-based estimate, supporting the use of the recombination-based approach as a proxy for the ionised, H$\alpha$-emitting mass.

\subsubsection{On the outflow mass and structure}

The significant difference between the volume-dependent estimate (i.e. $M_{\rm vol}$) and the volume-independent constraint (i.e. $M_{\rm ion}$) can, in turn, be used to evaluate key properties of the ejecta, such as $f_V$ (i.e. whether the outflow is clumpy) and the ionisation fraction $f_{\rm ion}$. In particular, by equating the two constraints on the mass involved in the nebular phase (see above), we obtain $M_{\rm vol}f_{\rm V} = M_{\rm ion}/f_{\rm ion}$, or equivalently, $M_{\rm ion}/M_{\rm vol} = f_{\rm V}f_{\rm ion} \sim 10^{-3}$. For this estimate, we used $M_{\rm vol} \sim 10^{-6} M_{\odot}$ and $M_{\rm ion} \sim 10^{-9} M_{\odot}$, corresponding to the case of $n_{\rm H} \sim 10^{11}$ cm$^{-3}$.

This low ratio between the two mass constraints suggests that both $f_{\rm V}$ and $f_{\rm ion}$ are significantly below unity. Clumpiness is further supported by the multiphase nature of the wind in V404~Cyg, where, in addition to the optical outflow discussed in this work, conspicuous signatures of X-ray winds are contemporaneously detected (\citealt{MunozDarias2022}; see also \citealt{King2015}). Clumping has also been inferred from modelling the optical spectra of the wind-bearing BHXB MAXI~J1820+070, for which $f_{\rm V} \sim 0.02$ was proposed \citep{Koljonen2023}. More generally, significant clumping has been invoked as a mechanism to mitigate over-ionisation in winds from accreting systems (e.g. \citealt{Matthews2016, Mosallanezhad2025}). All considered, values in the range $f_{\rm V} \sim 0.1$ to 0.01 are easy to reconcile with observations. This implies $f_{\rm ion} \sim f_{\rm V}$, consistent with a partially ionised ejecta. This estimate yields $M_{\rm out} = M_{\rm vol}f_{\rm V} = M_{\rm ion}/f_{\rm ion} \sim 10^{-7}-10^{-8}\,M_{\odot}$.
We note that recomputing our models with values down to $f_{\rm V} = 0.01$ produces negligible changes in the predicted line ratios, as the local physical conditions within individual clumps remain unchanged (e.g. \citealt{Osterbrock1959}).

Taking the $e$-folding timescale of 0.7~days to be representative of the evolution of the nebular phase, the previously estimated mass values correspond to the observable material during this interval. Extrapolating over the full duration of the nebular phase ($\sim4$~days) favours a total ejected mass of the order of $10^{-7}\,M_{\odot}$. This agrees with broad estimates derived purely from observational trends, such as the diffusion timescale of the nebular phase ($10^{-8}-10^{-5}\,M_{\odot}$; \MD). However, this is somewhat lower than the value derived from recombination estimates ($\sim4\times10^{-6}\,M_{\odot}$; \citealt{Casares2019}). 

Finally, we note that the mass involved in the nebular phase represents a fraction of the total mass launched during the outburst, as strong P-Cygni profiles were observed from the onset of activity and earlier ejecta no longer contribute to the nebular emission. Nevertheless, the inferred outflow mass is higher than both the total mass accreted during the 2015 outburst ($0.3$--$1.1\times10^{-8}\,M_{\odot}$) and the mass transferred during quiescence between the 1989 and 2015 outbursts ($3\times10^{-8}\,M_{\odot}$; \MD), further supporting the idea that at least in extreme cases such as V404~Cyg, winds have a strong impact on the black hole accretion process.

\section{Conclusions}
In this work, we presented the first modelling of the nebular phase of a BHXB. Our photoionisation simulations reproduce the key observables of V404~Cygni and show that a broad range of gas densities ($n_{\rm{H}}\sim10^{10.5}$–$10^{12.5}$\,cm$^{-3}$) and ionisation parameters ($-1.5 \lesssim \log\xi \lesssim 1$) are consistent with the observed nebular emission. We also provided new estimates of the outflow mass, finding that it likely exceeds the mass accreted during the outburst, consistent with previous observational studies. Finally, our results suggest that the nebular phase may represent a common evolutionary stage in black hole X-ray binaries, emerging as the ejected material cools and recombines following the decay of the X-ray luminosity. Therefore, these results strongly motivate spectroscopic monitoring of the final stages of future BHXB outbursts.

\begin{acknowledgements}
AA and TMD acknowledge support by the Spanish \textit{Agencia estatal de investigaci\'on} via PID2021-124879NB-I00 and PID2024-161863NB-I00.
JAFO acknowledges financial support by the Spanish Ministry of Science and Innovation (MCIN/AEI/10.13039/501100011033), by ``ERDF A way of making Europe'' and by ``European Union NextGenerationEU/PRTR'' through the grants PID2021-124918NB-C44 and CNS2023-145339; MCIN and the European Union -- NextGenerationEU through the Recovery and Resilience Facility project ICTS-MRR-2021-03-CEFCA. JC acknowledges support by the Spanish Ministry of Science via the Plan de Generaci\'on de Conocimiento through grant PID2022-143331NB-100. DMS acknowledges support via a Ramon y Cajal Fellowship RYC2023-044941, funded by MCIU/AEI/10.13039/501100011033 and FSE+. JHM acknowledges funding from a Royal Society University Research Fellowship (URF\textbackslash R1\textbackslash221062). This work is part of grant CEX2025-001609-S, awarded to the Instituto de Astrofísica de Canarias under the Severo Ochoa Centre of Excellence program and funded by MICIU/AEI/10.13039/501100011033. The authors wish to acknowledge the contribution of the IAC High-Performance Computing support team and hardware facilities to the results of this research.
\end{acknowledgements}

\bibliographystyle{aa} 
\bibliography{biblio} 

\begin{appendix} 

\begin{figure*} 
\section{BD and I$_{ratio}$}\label{appendice_BDIratio}
\centering
\includegraphics[width=\textwidth]{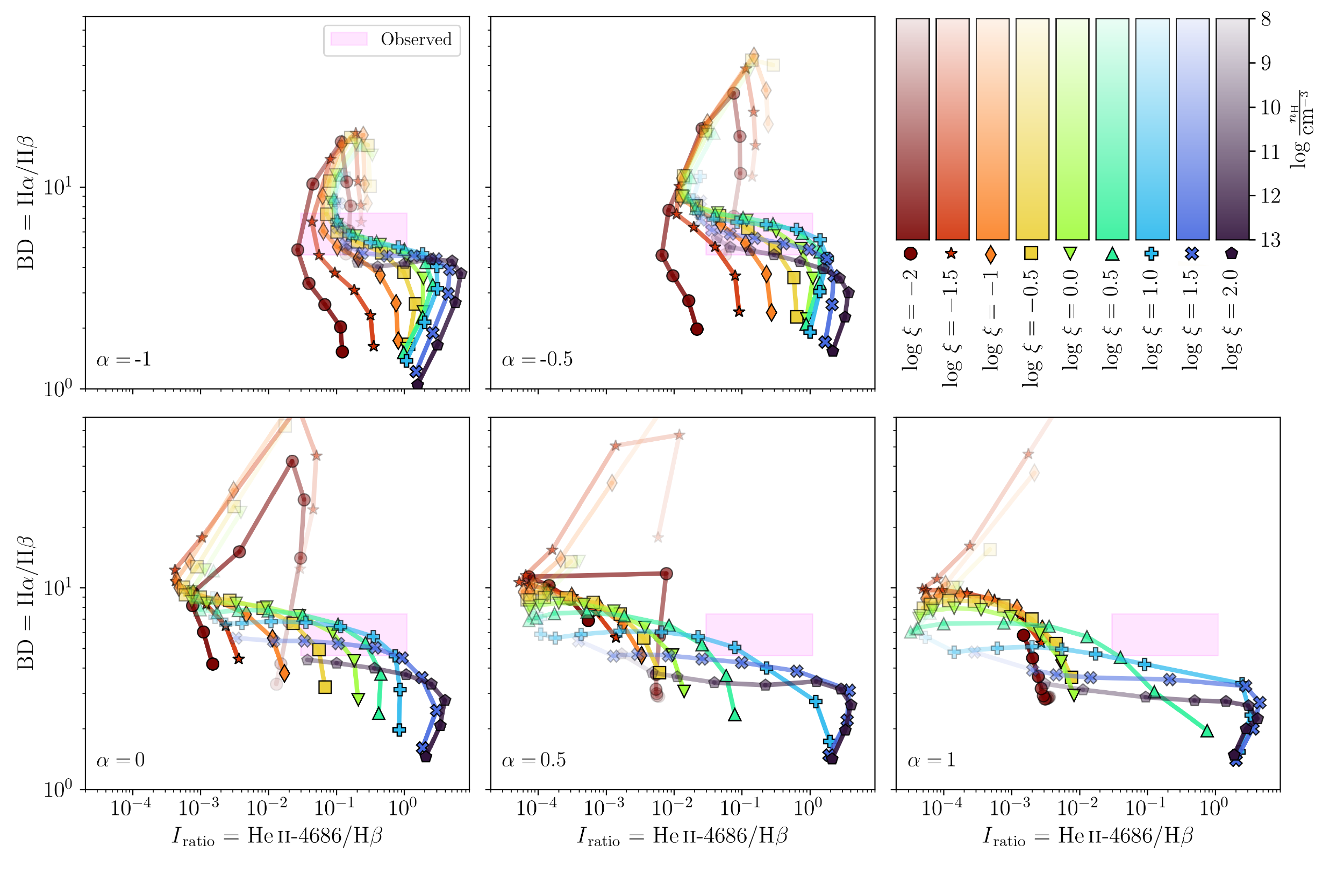}
\caption{BD vs I$_{\rm{ratio}}$ for the grid of models with fixed spectral index $\alpha$. From top to bottom and left to right, $\alpha=-1, \, -0.5,\, 0, \, 0.5, \, 1$. Models with $\log n_{\rm{H}} > 13$ are excluded from the figure.  Symbols and colours follow the conventions used in previous figures.}
\label{BD_Iratio_big}
\end{figure*}

\begin{figure*}
\section{BD and H$\alpha$ EW}\label{appendice_BDEW}
\centering
\includegraphics[width=\textwidth]{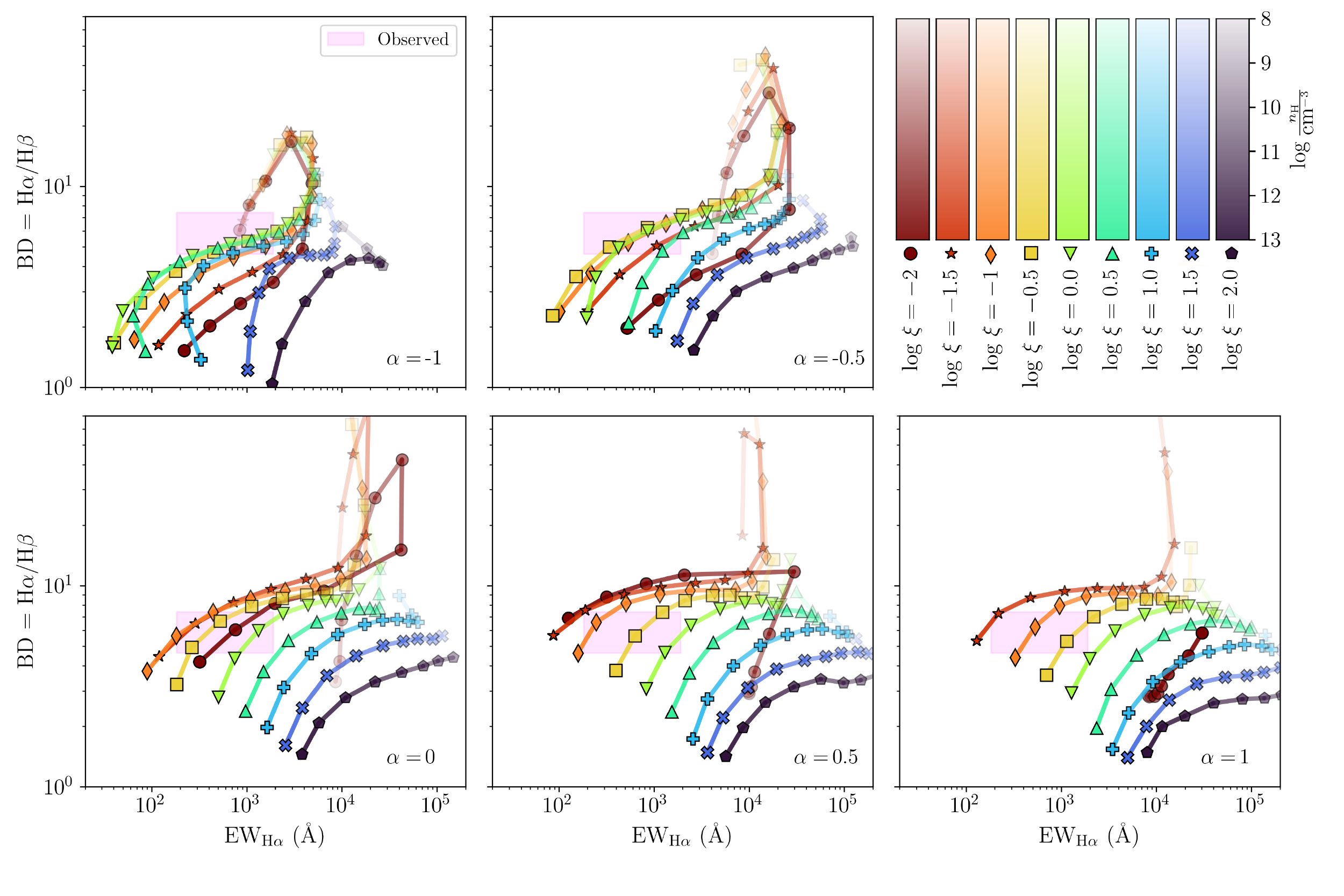}
\caption{BD vs H$\alpha$ EW for the grid of models with fixed spectral index $\alpha$. From top to bottom and left to right, $\alpha=-1, \, -0.5,\, 0, \, 0.5, \, 1$. Symbols and colours follow the conventions used in previous figures.}
\label{BD_EW_big}
\end{figure*}

\begin{figure*}
\section{Charts}\label{appendix_table}
\centering
\includegraphics[width=\textwidth]{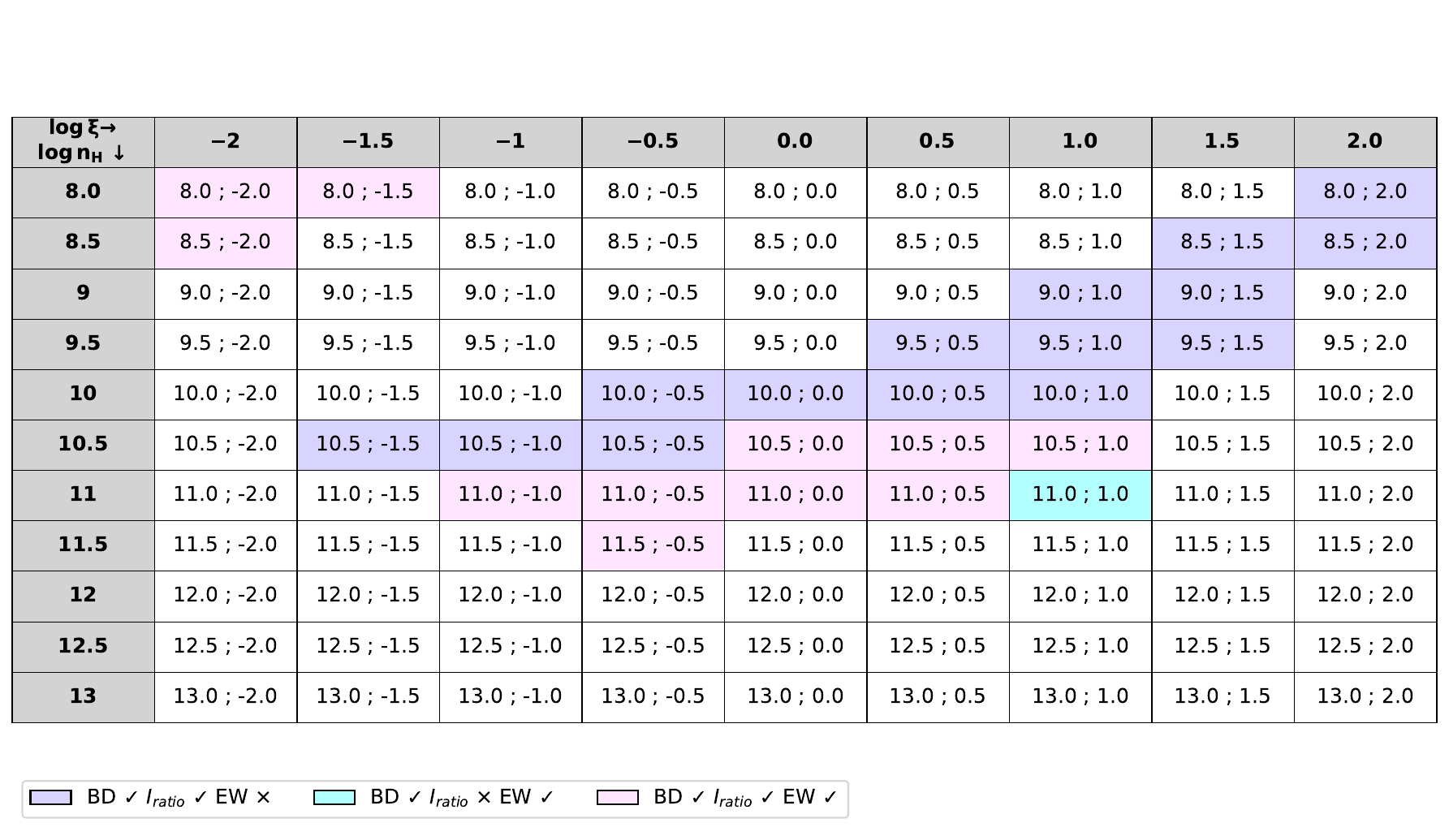}
\caption{Model grid for the spectral index $\alpha = -1$. The models highlighted in violet simultaneously reproduce the observed values of BD and $I_{\mathrm{ratio}}$; those in cyan reproduce BD and EW$_{\mathrm{H\alpha}}$; and those in pink reproduce all three quantities simultaneously. Models labelled as ``--'' indicate aborted simulations.}
\end{figure*}

\begin{figure*}
\centering
\includegraphics[width=\textwidth]{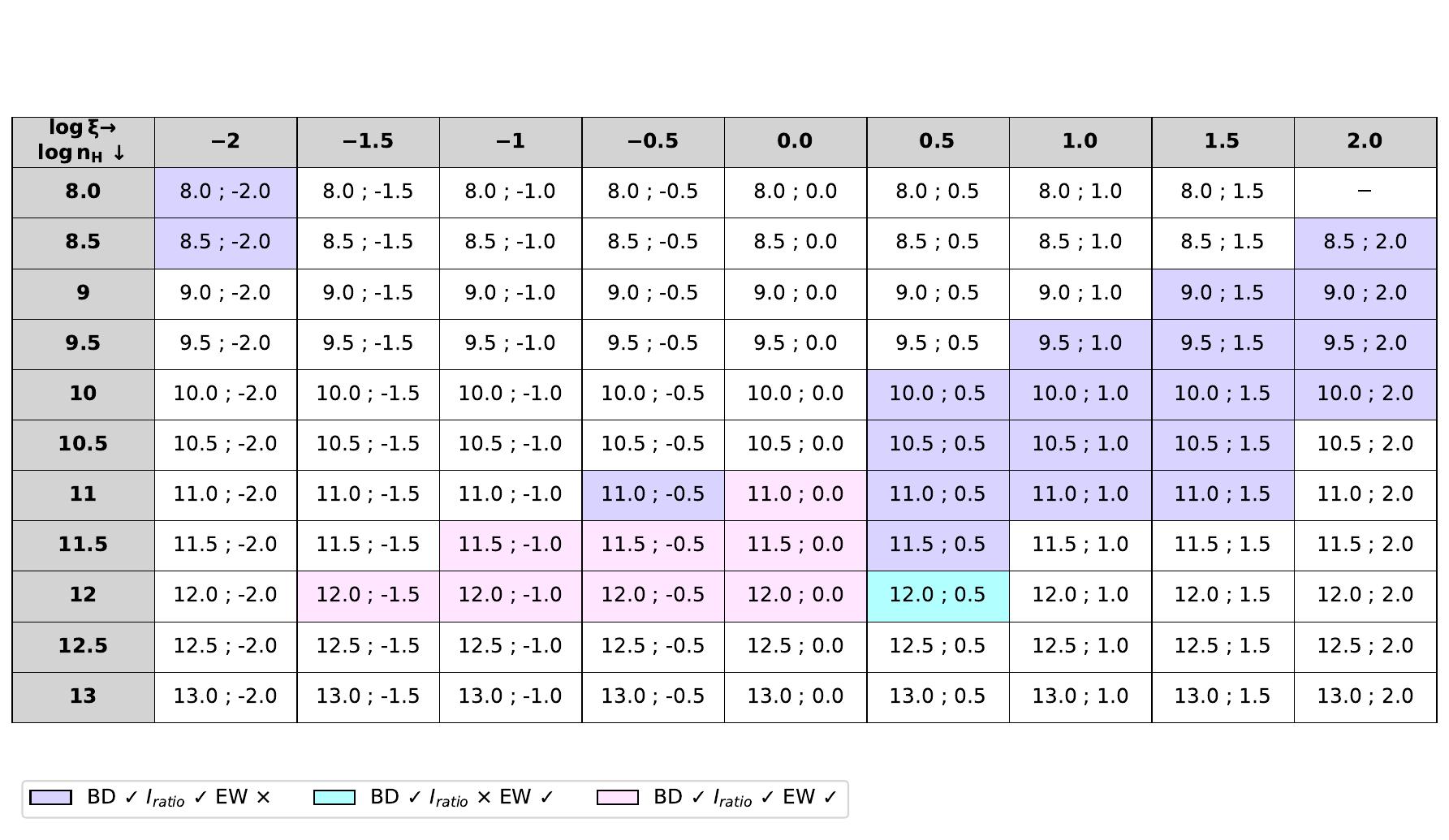}
\caption{Model grid for the spectral index $\alpha = -0.5$. Colours and symbols are the same as in the previous figure.}
\end{figure*}

\begin{figure*}
\centering
\includegraphics[width=\textwidth]{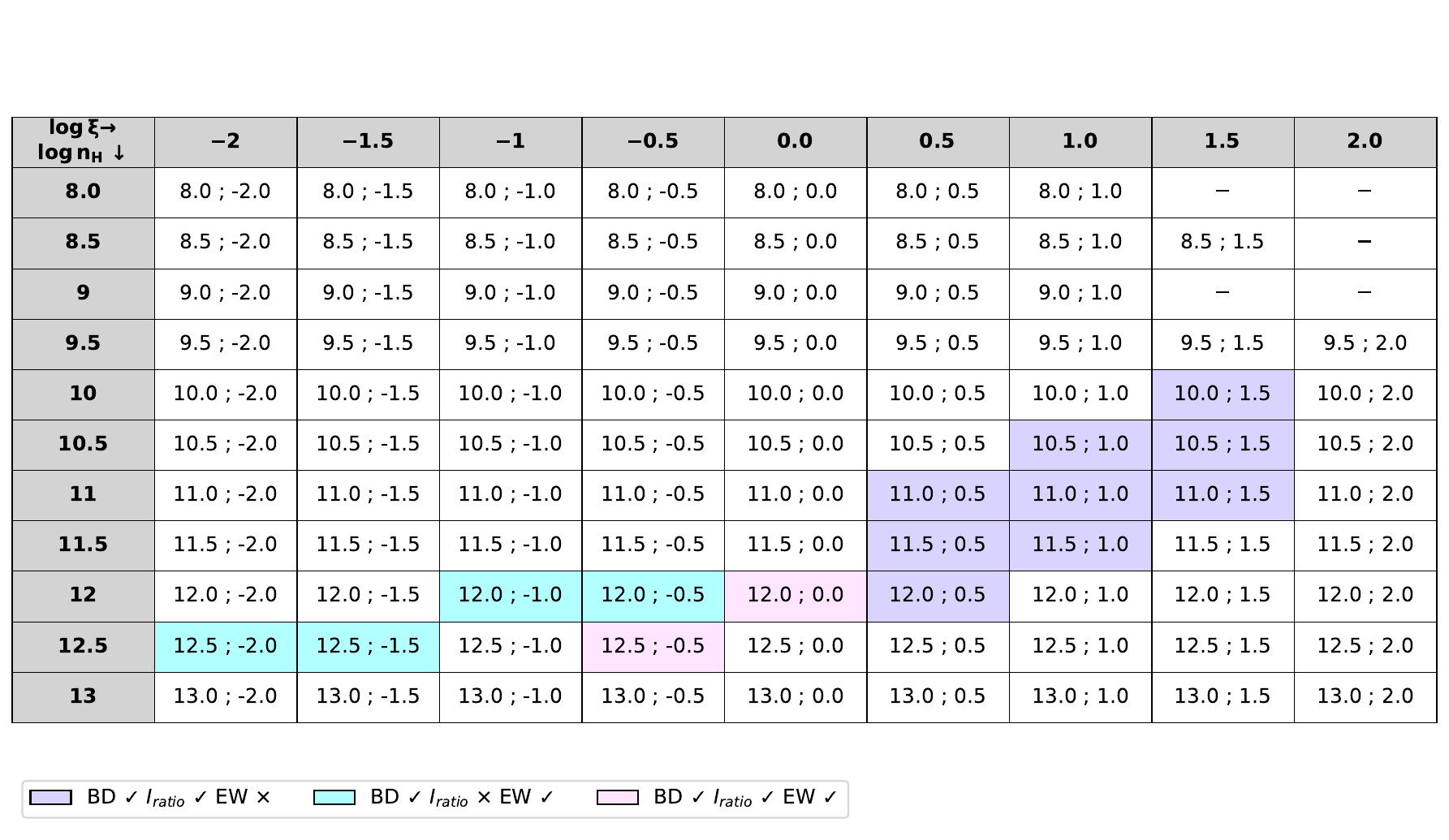}
\caption{Model grid for the spectral index $\alpha = 0$. Colours and symbols are the same as in the previous figure.}
\end{figure*}

\begin{figure*}
\centering
\includegraphics[width=\textwidth]{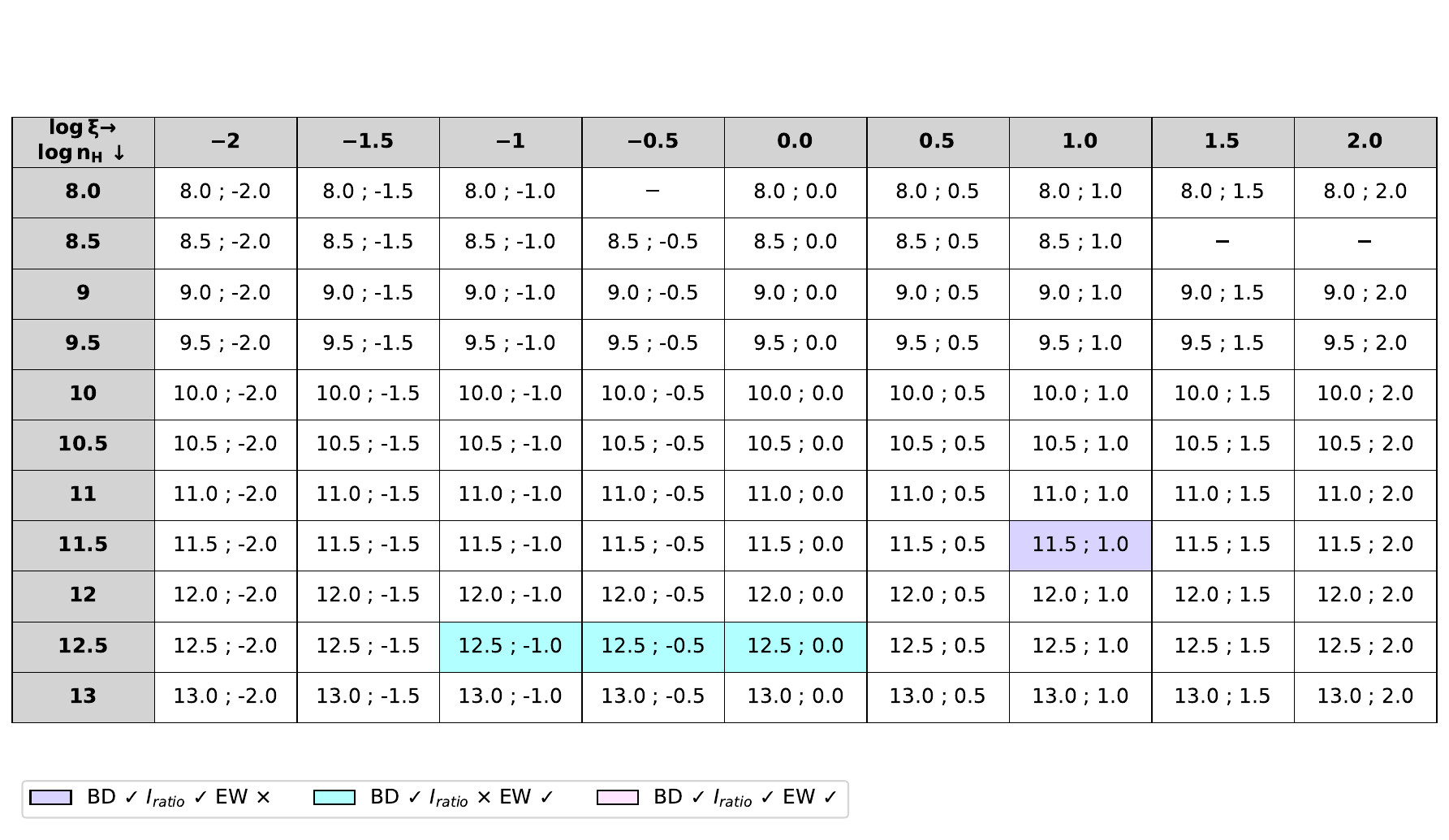}
\caption{Model grid for the spectral index $\alpha = 0.5$. Colours and symbols are the same as in the previous figure.}%
\end{figure*}

\begin{figure*}
\centering
\includegraphics[width=\textwidth]{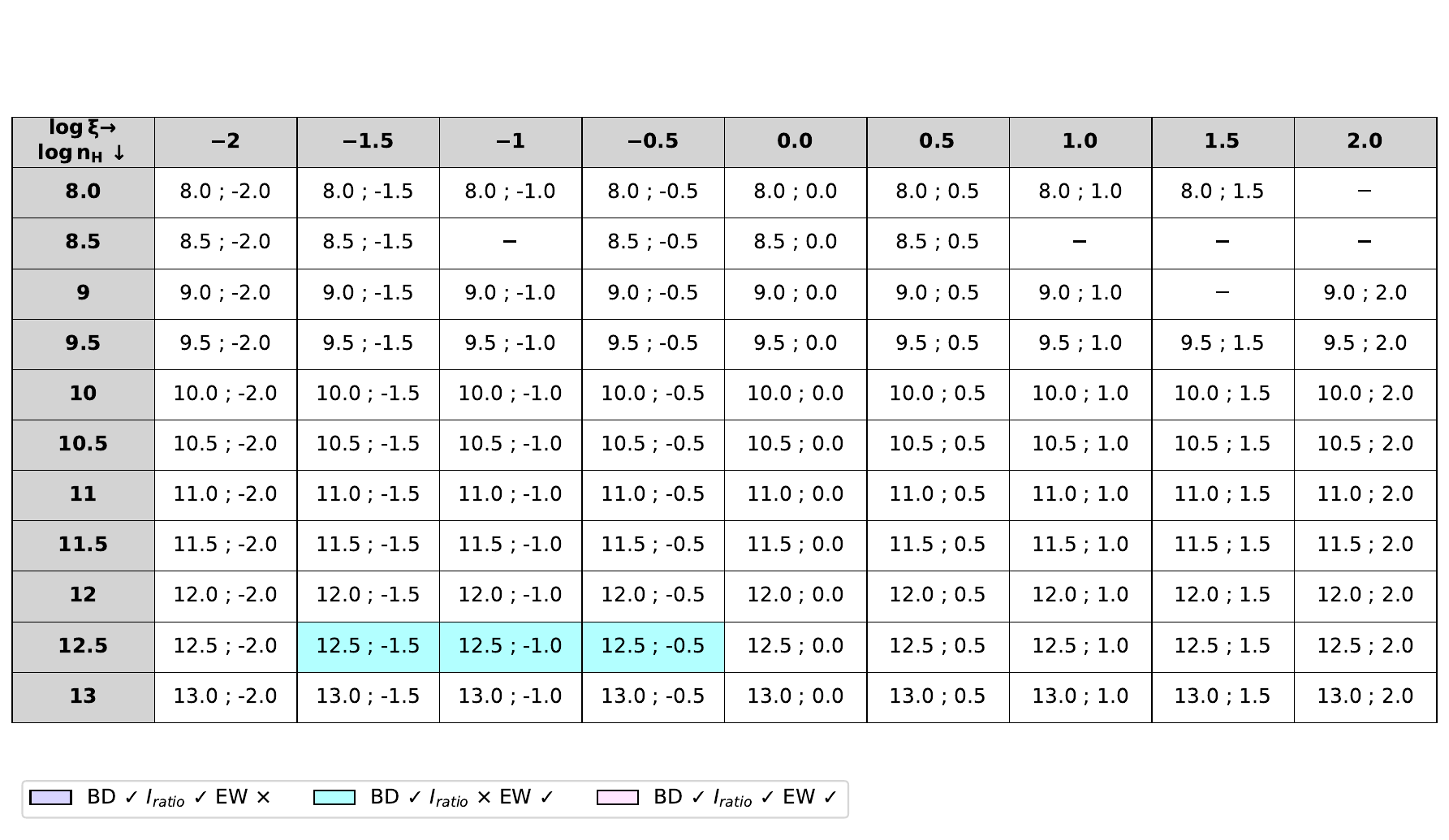}
\caption{Model grid for the spectral index $\alpha = 1$. Colours and symbols are the same as in the previous figure.}
\end{figure*}

\begin{figure*}
\section{Outflow mass}\label{appendice_mass}
\centering
\includegraphics[width=\textwidth]{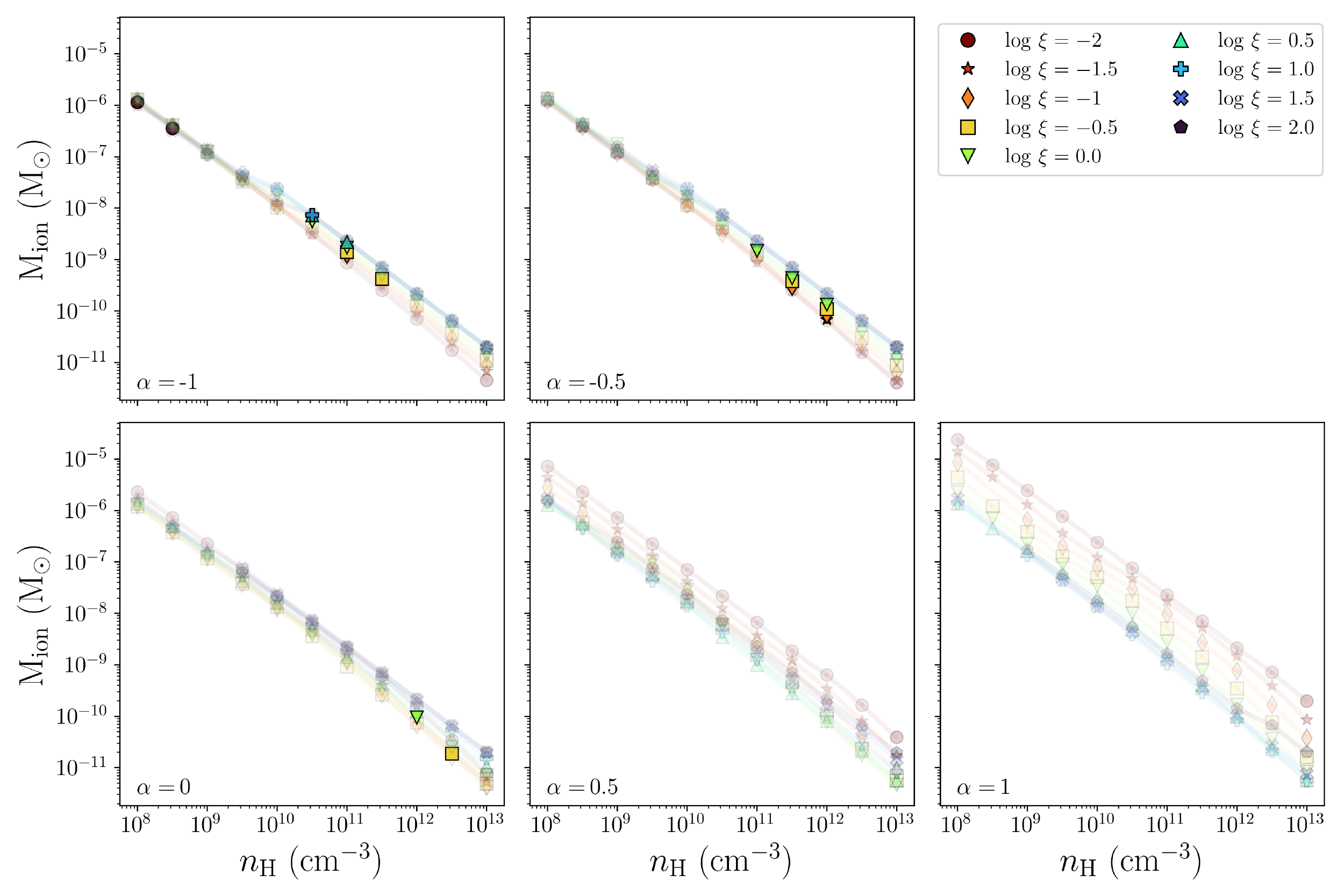}
\caption{Ionised mass as a function of hydrogen density, computed using Eq.~\ref{Mion2}. Each subplot shows a subset of models with different spectral index $\alpha$. Models with the same ionisation parameter $\xi$ are shown using identical symbols and colours and are connected by lines. Solid coloured symbols indicate models that reproduce the observed values of BD, $I_{ ratio}$, and H$\alpha$ equivalent width.}
\label{Mass_big}
\end{figure*}

\clearpage 

\section{Outflow mass from emissivity}\label{App:mass_from_emissivity}

A lower limit to the outflow mass can also be obtained directly from the H$\alpha$ emissivities predicted by our \textsc{Cloudy} models, complementing the recombination-based approach of Sect.~\ref{recomb_estimate}. We write the H$\alpha$ luminosity as:
\begin{equation}
  L_\mathrm{H\alpha} = \int_V j_{H\alpha}(r)\, dV \approx j_{H\alpha}\, V = 4 \pi d^2 F_\mathrm{H\alpha},
\end{equation}
where $j_{\mathrm{H}\alpha}$ is the volume emissivity, and substitute $V$ into $M_\mathrm{emis} = 1.4\,n_\text{H} m_\mathrm{p} V$, where $M_\mathrm{emis}$ is the mass contributing to the H$\alpha$ emission. This yields:
\begin{equation}\label{M_emissivity}    M_\text{emis} \sim 1.4 n_{\rm{H}} m_\text{p} \frac{4\pi d^2 F_\mathrm{H\alpha}}{j_{H\alpha}}.
\end{equation}

\begin{figure*}
\centering
\includegraphics[width=\textwidth]{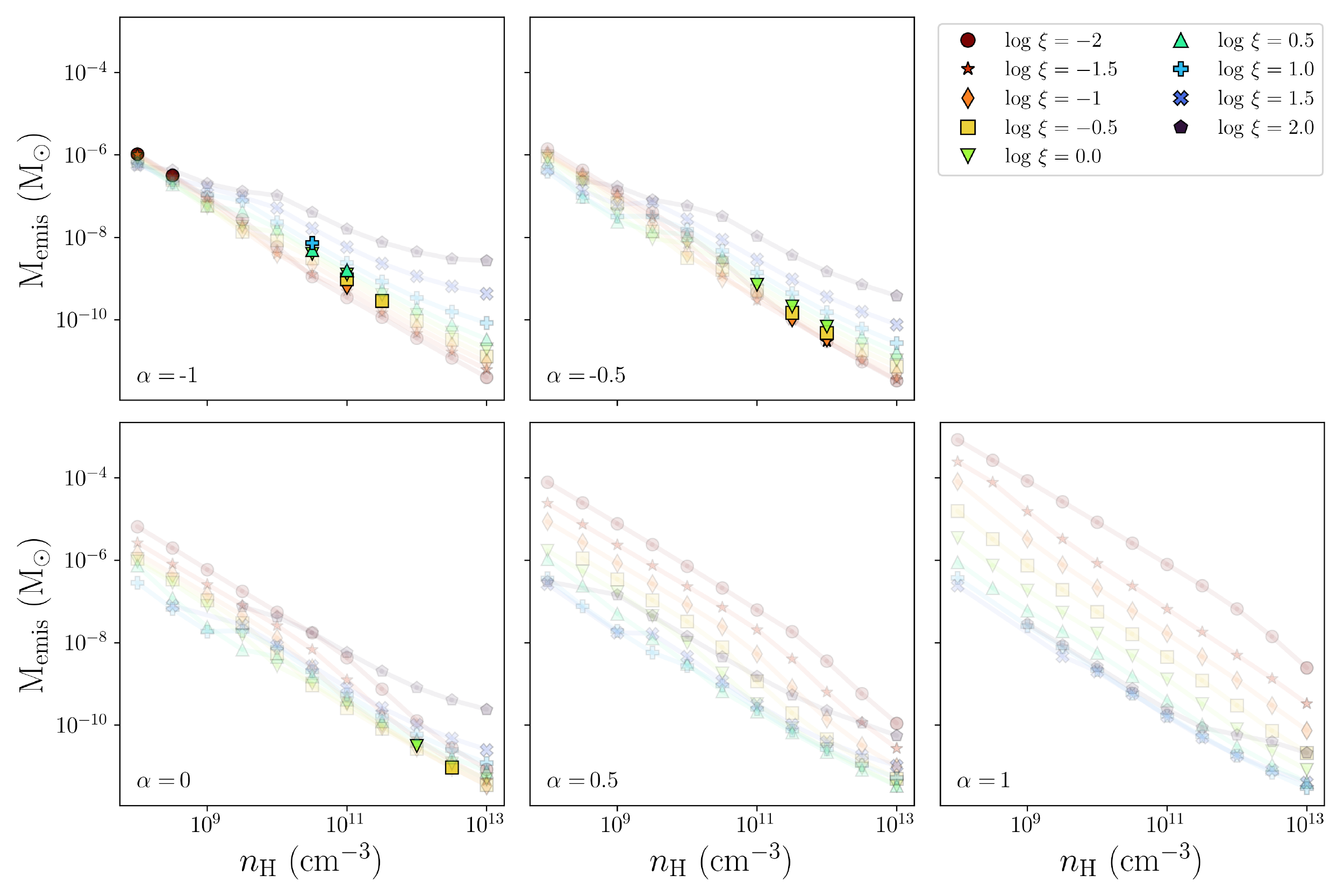}
\caption{Mass contributing to the H$\alpha$ emission as a function of hydrogen density, computed using Eq.~\ref{M_emissivity}. Each subplot shows a subset of models with different spectral index $\alpha$. Models with the same ionisation parameter $\xi$ are shown using identical symbols and colours and are connected by lines. Solid coloured symbols indicate models that reproduce the observed values of BD, $I_{ ratio}$, and H$\alpha$ equivalent width.}
\label{Fig:mass_emissivity}
\end{figure*}

We adopt $F_{\mathrm{H}\alpha}$ equal to the maximum H$\alpha$ flux observed during the nebular phase, and $j_{\mathrm{H}\alpha}$ equal to the maximum H$\alpha$ emissivity predicted by each model. This choice provides a lower limit on the outflow mass. In fact, employing the  maximum emissivity as representative for the entire emitting volume reduces the mass required to account for the observed flux. The results for the full grid is presented in Figure~\ref{Fig:mass_emissivity}. 
For the models reproducing the observables, we obtain $M_{\rm emis}\sim9 \times 10^{-12}-7 \times 10^{-9}\,M_{\odot}$, comparable to the recombination-based estimate $M_{\rm ion}$ (Section \ref{recomb_estimate}).

The discrepancy between the recombination-based and emissivity-based mass estimates stems from the assumption in the former of pure recombination being the only mechanism contributing to the H$\alpha$ line emission. However, at the high densities ($\log n_\mathrm{e} \gtrsim 8 \text{ cm}^{-3}$) considered in our modelling, Balmer self-absorption and collisional de-excitation are expected to contribute to the emission \citep{Netzer1975, Drake1980}. The local escape probability formalism implemented in \textsc{Cloudy} for the computation of line emissivities provides a framework to account for these effects.  The order-of-magnitude agreement between the two methods nevertheless confirms that the recombination-based approach of Sect.~\ref{recomb_estimate} provides a reliable and practical estimate of the mass contributing to H$\alpha$ emission, particularly useful for observational studies where detailed photoionisation modelling is unavailable.

\end{appendix}

\end{document}